\RequirePackage{xcolor}
\documentclass[journal,twoside,web]{ieeecolor}

\let\labelindent\relax
\usepackage[table]{xcolor}
\usepackage{orcidlink}
\usepackage{booktabs}
\usepackage{cite}
\usepackage{amsmath,amssymb,amsfonts}
\usepackage[nolist,nohyperlinks]{preamble/acronym}
\usepackage[inline]{enumitem}
\usepackage[detect-weight=true, detect-family=true, per-mode=fraction]{siunitx}
\usepackage{algpseudocode}
\usepackage{algorithm}
\usepackage{svg}
\usepackage{multirow}
\usepackage{balance}
\usepackage{graphicx}
\usepackage{tikz}
\usetikzlibrary{automata, positioning, arrows}
\usepackage[nameinlink, capitalise]{cleveref}
\usepackage[usestackEOL]{stackengine}
\usepackage{dblfloatfix} 

\DeclareSIUnit\ohm{\ensuremath\Omega}
\DeclareSIUnit\db{dB}
\DeclareSIUnit{\belmilliwatt}{Bm}
\DeclareSIUnit{\belisotrope}{Bi}
\DeclareSIUnit{\bel}{B}
\DeclareSIUnit{\bitpersecond}{bps}
\DeclareSIUnit{\samplepersecond}{Sps}
\DeclareSIUnit{\nothing}{\relax}
\usepackage[export]{adjustbox}

\usepackage{pifont}
\crefformat{footnote}{#2\footnotemark[#1]#3}
\makeatletter
\newcommand*{\org@overidelabel}{}
\let\org@overridelabel\@verridelabel
\@ifpackagelater{acronym}{2015/03/21}{
  \renewcommand*{\@verridelabel}[1]{%
    \@bsphack
    \protected@write\@auxout{}{\string\AC@undonewlabel{#1@cref}}%
    \org@overridelabel{#1}%
    \@esphack
  }%
}{
  \renewcommand*{\@verridelabel}[1]{%
    \@bsphack
    \protected@write\@auxout{}{\string\undonewlabel{#1@cref}}%
    \org@overridelabel{#1}%
    \@esphack
  }%
}

\newcommand{\linebreakand}{%
  \end{@IEEEauthorhalign}
  \hfill\mbox{}\par
  \mbox{}\hfill\begin{@IEEEauthorhalign}
}
\makeatother

\usepackage{wrapfig}
\usepackage{jsen}

\usepackage{eso-pic}
\usepackage{url}
\AddToShipoutPictureBG*{
  \AtPageUpperLeft{%
    \put(0,-30){\raisebox{15pt}{\makebox[\paperwidth]{\begin{minipage}{21cm}\centering
      \textcolor{gray}{This work has been accepted for publication in the IEEE Sensors Journal.\\DOI: \href{https://doi.org/10.1109/JSEN.2026.3652283}{10.1109/JSEN.2026.3652283}} 
    \end{minipage}}}}%
  }
  \AtPageLowerLeft{%
    \raisebox{20pt}{\makebox[\paperwidth]{\begin{minipage}{21cm}\centering
      \textcolor{gray}{ \copyright 2026 IEEE.  Personal use of this material is permitted.  Permission from IEEE must be obtained for all other uses, in any current or future media, including reprinting/republishing this material for advertising or promotional purposes, creating new collective works, for resale or redistribution to servers or lists, or reuse of any copyrighted component of this work in other works.
      }
    \end{minipage}}}%
  }
}

\def\BibTeX{{\rm B\kern-.05em{\sc i\kern-.025em b}\kern-.08em
    T\kern-.1667em\lower.7ex\hbox{E}\kern-.125emX}}
\definecolor{abstractbg}{rgb}{0.89804,0.94510,0.83137}

\begin{acronym}
    \acro{A-TDOA}{asynchronous \acl{TDOA}}
    \acro{ADS-TWR}{alternative \acl{DS-TWR}}
    \acro{AIMD}{additive-increase/multiplicative-decrease}
    \acro{AMCL}{Adaptive Monte Carlo Localization}
    \acro{AP-TWR}{active-passive \acl{TWR}}
    \acro{ASK}{amplitude shift-keying}
    \acro{AoA}{angle of arrival}
    \acro{BLE}{Bluetooth Low Energy}
    \acro{CA}{cumulative average}
    \acro{CC-SS-TWR}{\acl{CFO} corrected \acl{SS-TWR}}
    \acro{CFO}{carrier frequency offset}
    \acro{CIR}{channel-impulse response}
    \acro{CPU}{central processing unit}
    \acro{DL-TDOA}{downlink \acl{TDOA}}
    \acro{DL}{deep learning}
    \acro{DS-TWR}{double-sided \acl{TWR}}
    \acro{DWT}{Data Watchpoint and Trace}
    \acro{EKF}{Extended Kalman Filter}
    \acro{ESA}{European Space Agency}
    \acro{FFT}{fast Fourier transform}
    \acro{FHC}{finite horizon control}
    \acro{FPU}{floating point unit}
    \acro{FSM}{finite state machine}
    \acro{GDOP}{geometric dilution of precision}
    \acro{GNSS}{global navigation satellite system}
    \acro{GPIO}{general purpose input-output}
    \acro{GT}{ground truth}
    \acro{HDR}{high data rate}
    \acro{IC}{integrated circuit}
    \acro{IMU}{inertial measurement unit}
    \acro{IQR}{interquartile range}
    \acro{IoT}{Internet of Things}
    \acro{KF}{Kalman Filter}
    \acro{LDR}{low data rate}
    \acro{LM}{Levenberg–Marquardt}
    \acro{LNA}{low-noise amplifier}
    \acro{LOS}{line-of-sight}
    \acro{LTE-M}{Long-Term Evolution Machine Type Communication}
    \acro{LiDAR}{light detection and ranging}
    \acro{LiPo}{lithium-polymer}
    \acro{MCPD}{multi-carrier phase-difference}
    \acro{MCU}{microcontroller}
    \acro{ML}{machine learning}
    \acro{MPC}{model predictive control}
    \acro{MPPT}{maximum power-point tracking}
    \acro{MRI}{Magnetic Resonance Imaging}
    \acro{NI}{non parametric information}
    \acro{NLOS}{non-line-of-sight}
    \acro{OOK}{on-off keying}
    \acro{PCB}{printed circuit board}
    \acro{PC}{personal computer}
    \acro{PDA}{personal digital assistant}
    \acro{PF}{Particle Filter}
    \acro{PLL}{phase-locked loop}
    \acro{QoS}{quality of service}
    \acro{RAM}{random-access memory}
    \acro{RFID}{radio frequency identification}
    \acro{RF}{radio frequency}
    \acro{RSSI}{received signal strength indicator}
    \acro{RTLS}{real-time locating system}
    \acro{RTT}{round-trip time}
    \acro{SPI}{serial peripheral interface}
    \acro{SS-TWR}{single-sided \acl{TWR}}
    \acro{SoC}{state-of-charge}
    \acro{TDMA}{time-division multiple-access}
    \acro{TDOA}{time difference of arrival}
    \acro{TWR}{two-way ranging}
    \acro{ToF}{time-of-flight}
    \acro{UAA}{uniform angular array}
    \acro{UAV}{unmanned aerial vehicle}
    \acro{USB}{universal serial bus}
    \acro{UWB}{ultra-wideband}
    \acro{WuC}{wake-up call}
    \acro{WuR}{wake-up radio}
\end{acronym}

\begin{document}
\title{Eco-WakeLoc: An Energy-Neutral and Cooperative \acs{UWB} Real-Time Locating System}
\author{\hspace{-0.02cm}Silvano Cortesi\orcidlink{0000-0002-2642-0797}, \textit{Graduate Student Member, IEEE}, Lukas Schulthess\orcidlink{0000-0002-6027-2927}, \textit{Graduate Student Member, IEEE},\\ Davide Plozza\orcidlink{0009-0002-1197-951X}, \textit{Graduate Student Member, IEEE}, Christian Vogt\orcidlink{0000-0003-4551-4876}, \textit{Member, IEEE}, and\\ Michele Magno\orcidlink{0000-0003-0368-8923}, \textit{Senior Member, IEEE}
\thanks{\textit{(Corresponding author: Silvano Cortesi)}}
\thanks{Silvano Cortesi, Lukas Schulthess, Davide Plozza, Christian Vogt and Michele Magno are with the Center for Project-Based Learning, ETH Zürich, 8092 Zürich, Switzerland (e-mail: firstname.lastname@pbl.ee.ethz.ch)}}
\IEEEtitleabstractindextext{
\fcolorbox{abstractbg}{abstractbg}{
\begin{minipage}{\textwidth}
\begin{wrapfigure}[12]{c}{3.25in}
\includesvg[width=3.18in]{figures/abstract-graphic_nofonts.svg}
\end{wrapfigure}
\begin{abstract}
Indoor localization systems face a fundamental trade-off between energy efficiency and responsiveness, which is especially important for emerging use cases such as mobile robots operating in GPS-denied environments. Traditional \ac{RTLS} either require continuously powered infrastructure, limiting their scalability, or are limited by their responsiveness. This work presents Eco-WakeLoc, designed to achieve centimeter-level \ac{UWB} localization while remaining energy-neutral by combining ultra-low power \acp{WuR} with solar energy harvesting. By activating anchor nodes only on demand, the proposed system eliminates constant energy consumption while achieving centimeter-level positioning accuracy.

To reduce coordination overhead and improve scalability, Eco-WakeLoc employs cooperative localization where \textit{active} tags initiate ranging exchanges, while \textit{passive} tags opportunistically reuse these messages for \ac{TDOA} positioning. An \ac{AIMD}-based energy-aware scheduler adapts localization rates according to the harvested energy, thereby maximizing the overall performance of the sensor network while ensuring long-term energy neutrality.

A comprehensive evaluation demonstrates centimeter-level accuracy with average errors in 3D space of \qty{21.89}{\centi\meter} for \textit{active} tags using Levenberg-Marquardt trilateration, and \qty{25.70}{\centi\meter}  for \textit{passive} tags through cooperative \ac{TDOA}. The measured energy consumption is only \qty{3.22}{\milli\joule} per localization for \textit{active} tags, \qty{951}{\micro\joule} for \textit{passive} tags, and \qty{353}{\micro\joule} for anchors. Real-world deployment on a quadruped robot with nine anchors confirms the practical feasibility, achieving an average accuracy of \qty{43}{\centi\meter} in dynamic indoor environments. Year-long simulations starting from an empty battery, and using solar harvesting from a \qty{10}{\centi\meter\squared} cell under typical indoor lighting conditions, show that tags achieve an average of \num{2031} localizations per day, while retaining over \qty{7}{\percent} battery capacity after one year -- demonstrating that the \ac{RTLS} achieves sustained energy-neutral operation.

Eco-WakeLoc demonstrates that high-accuracy indoor localization can be achieved at scale without continuous infrastructure operation, combining energy neutrality, cooperative positioning, and adaptive scheduling to enable maintenance-free deployments suitable for large-scale \acl{IoT} applications.
\end{abstract}

\end{minipage}}}
\maketitle

\acresetall
\section{Introduction}\label{sec:introduction}
Indoor localization is becoming a cornerstone for a wide range of applications, from smart buildings~\cite{chen24_centim_level_indoor_posit_with} and asset tracking~\cite{silva22_real_world_deploy_low_cost} to robotics~\cite{bouazzaoui22_enhan_rgb_d_slam_perfor} and pervasive \ac{IoT} service~\cite{hayward22_survey_indoor_locat_techn_techn_applic_indus}. Overall, these applications require precise, on-demand, and scalable position tracking, while allowing for extended or even self-sustainable operation of the object localizing itself (the tag), as well as the infrastructure (the anchors)~\cite{hayward22_survey_indoor_locat_techn_techn_applic_indus, venkatapathy15_phynode}. To achieve these goals, advances in localization technologies, ultra-low-power hardware, efficient localization algorithms, and energy-efficient localization scheduling must be combined to create solid sensing systems for \acp{RTLS}~\cite{silva22_real_world_deploy_low_cost, mayer24_self_sustain_ultraw_posit_system, chen24_room_level_indoor_local_using}.

Concerning localization technology, out of the numerous wireless technologies for indoor localization, \ac{UWB} has emerged as a leading choice due to its capability for centimeter-level localization~\cite{qi24_calib_compen_anchor_posit_uwb_indoor_local,sayfoori25_high_precis_uwb_sensor_based,yang22_high_precis_uwb_based_local} and its high resilience to interference~\cite{li25_indoor_uwb_local_method_based,bilge22_evaluat_ultra_wide_band_techn}. A typical \ac{UWB}-based \ac{RTLS} consists of static anchors with known positions and mobile tags whose locations are estimated using techniques such as \ac{TDOA} or multilateration~\cite{patru23_flext, laadung22_novel_activ_passiv_two_way}.

Such extensive infrastructure, with anchors requiring wired power connections~\cite{laadung22_novel_activ_passiv_two_way, patru23_flext}, hinders scalability and impedes economical retrofitting of existing buildings~\cite{sullivan23_total_cost_owner_real_time}. As already demonstrated commercially for \ac{BLE} beacon-based indoor navigation, retrofitting is adapted more easily when the anchors or beacons are battery-powered and can operate for multiple years~\cite{polymaps}. Therefore, achieving ultra-low power operation is critical for widespread adoption of precise \acp{RTLS}, ideally allowing self-sustainable~\cite{ahmed19_optim_power_manag_with_guaran} operation to lower installation and maintenance costs. Thus, a key research challenge in realizing practical and large-scale deployments lies in reducing the power consumption of the entire \ac{RTLS} while maintaining high localization performance~\cite {boulos12_real_time_locat_system_rtls_healt, sullivan23_total_cost_owner_real_time}.

However, conventional radios used for localization, including \ac{UWB}, are highly inefficient during idle listening, since the transceiver consumes nearly the same amount of power while waiting for potential packets as it does during active communication~\cite{istomin21_janus,polonelli22_perfor_compar_decaw_dw100_dw300}. This constant energy drain becomes a significant bottleneck in battery-operated mobile systems. To address this issue, ultra-low-power \acp{WuR}~\cite{villani24_ultra_wideb_wake_up_receiv,kazdaridis21_a_novel_archit}, combined with emerging energy-aware algorithms, enable event-driven communication: devices remain in deep sleep, consuming only micro-watts, and activate the main transceiver only upon reception of a dedicated wake-up signal~\cite{verma22_novel_rf_energ_harves_event}.


This approach avoids the fundamental trade-off in traditional \ac{RTLS} solutions between energy efficiency and responsiveness: Duty cycling or time-scheduled communication schemes can significantly reduce power consumption~\cite{luder25_anitr,zhao21_uloc}, but they inevitably increase latency, limiting their use in time-critical and event-driven applications~\cite{mayer24_self_sustain_ultraw_posit_system}. Furthermore, time-scheduling approaches do not account for actual localization needs: when no assets need to be tracked, anchors are still activated regularly to ensure localization, resulting in unnecessary energy consumption~\cite{cortesi25_wakel}.
Conversely, continuously active devices allow for low-latency localization; however, they substantially limit the battery lifetime due to the increased energy cost. 

While \ac{WuR} hardware can provide the foundation for on-demand operation, it is the addition of intelligent, energy-aware algorithms that enables a system to strategically manage its resources to maintain long-term energy neutrality~\cite{balsamo16_hiber,giouroukis20}. 
Previous works have shown the theoretical promise of combining \ac{UWB} and \acp{WuR} for achieving accurate and energy-efficient localization, by operating the energy-consuming \ac{UWB} distance sensing in an event-driven fashion \textsc{WakeLoc}~\cite{cortesi25_wakel}, and is detailed as background in \cref{subsec:accuracy-algorithms}. In particular, \textsc{WakeLoc} focused on the localization protocol, while \textsc{EcoTrack}~\cite{giordano23_energ_aware_adapt_sampl_self} addressed the energy-aware scheduling. By dynamically adapting the sensing and communication rate to the harvested energy, \textsc{EcoTrack} ensures that devices remain energy-neutral while providing the highest possible amount of \ac{GNSS} localizations, thereby balancing performance and battery lifetime in an energy-harvesting deployment. 

 This work introduces a cooperative and energy-neutral \ac{RTLS} that integrates ultra-low power \acp{WuR} with \ac{UWB} ranging, addressing the open challenge of enabling sustainable high-accuracy localization despite \ac{UWB}’s inherently high energy consumption. Unlike our earlier studies -- \textsc{WakeLoc}, which focused on the localization protocol, and \textsc{EcoTrack}, which developed adaptive energy-aware scheduling -- the proposed approach combines both advances into a unified solution for the more demanding context of \ac{UWB}-based localization. By fusing ultra-low power wake-up signaling with high-precision \ac{UWB} ranging, tags and anchors remain in deep sleep until explicitly triggered, while still offering centimeter-level localization accuracy and communication on demand. The system is based on a newly developed sensor node that integrates \textsc{WakeMod}~\cite{schulthess25_wakem}, together with a \textsc{DWM3000} \ac{UWB} transceiver from \textsc{Qorvo} and solar energy harvesting capabilities. 


In addition, we develop a complete implementation platform that not only realizes the proposed system in hardware but also provides the foundation for experimental validation of its communication primitives and algorithmic intelligence. This platform serves as a proof of concept for sustainable localization, bridging theoretical design with real-world feasibility. The addition of indoor solar harvesting extends this paradigm further by enabling perpetual operation under realistic illumination conditions, thereby reducing maintenance needs and facilitating scalable, long-term deployments. The main contributions of this work are as follows:

\begin{enumerate}
  \item We introduce and evaluate an energy-aware scheduling strategy, aiming to maximize the localization rate of \textit{active} tags by dynamically scheduling ranging opportunities based on energy availability.
  \item A comprehensive characterization of the system's accuracy across multiple localization solvers and environments, identifying performance trade-offs relevant to real-world deployments.
  \item To demonstrate system functionality, an open-source\footnote{\label{github}\url{https://github.com/ETH-PBL/WakeLoc}} ultra-low-power sensor node was developed that integrates \textsc{WakeMod}, the \textsc{DWM3000} \ac{UWB} transceiver, and solar harvesting. A detailed evaluation of its power consumption and computational cost validates the feasibility of event-driven, high-accuracy localization with self-sustainable operation.
  \item A real-world deployment utilizing a quadruped robot as a mobile tag, as well as nine anchors, demonstrating practicality in dynamic indoor environments with a localization latency below \qty{83}{\milli\second}, as well as an open-source simulation framework\cref{github} showing the effects of large-scale deployments.
\end{enumerate}
\section{Related Works}\label{sec:related_works}
Developing an energy-neutral indoor localization system requires addressing challenges across multiple domains. The following discussion covers \ac{UWB}-based localization approaches and their scalability limitations, \ac{WuR} technologies enabling ultra-low-power operation, and positioning algorithms with their computational trade-offs.

\subsection{UWB-Based Indoor Localization Systems}
Most three-dimensional localization schemes are based on the capability of \ac{UWB} to accurately measure distance within centimeter precision~\cite{dotlic18_rangin_method_utiliz_carrier_frequen_offset_estim,coppens22_overv_uwb_stand_organ_ieee}. Multiple methods to exploit these measurements exist, an overview is presented in this section. 

Conventional \ac{TWR} approaches require continuous anchor operation to maintain low-latency localization functionality~\cite{cortesi25_wakel}. Each tag has to exchange multiple messages with at least four anchors, which significantly limits the system's scalability when multiple tags are trying to localize simultaneously~\cite{ridolfi18_analy_scalab_uwb_indoor_local,ramesh20_robus_scalab_techn_twr_tdoa}. Recent solutions addressed this through scheduling algorithms~\cite{yang22_vuloc}, anchor selection strategies~\cite{herbruggen24_real_time_anchor_node_selec}, and concurrent transmission techniques~\cite{corbalan20_ultra_wideb_concur_rangin,grosswindhager19_snapl}. While these approaches improve scalability, the fundamental energy problem of continuous anchor operation remains unsolved.

\ac{TDOA}-based systems offer better scalability by enabling multiple tags to localize simultaneously from synchronized anchor transmissions~\cite{santoro21_scale,patru23_flext}. The \ac{DL-TDOA} approach allows unlimited tag scaling since packet overhead and the resulting complexity of the non-linear systems of equations depend only on the anchor count~\cite{spirito98_hyperbolic_posi,gust03positioning_tdoa}. However, these systems require complex synchronization infrastructure and coordination between anchors, which limits practical deployment flexibility and introduces scalability issues related to the number of anchors.

Recent hybrid approaches have attempted to combine the benefits of both paradigms. Some systems combine \ac{TWR} and \ac{TDOA} to achieve higher accuracy~\cite{kolakowski16_tdoa_twr_uwb}, or to facilitate anchor synchronization~\cite{ramesh20_robus_scalab_techn_twr_tdoa}, while others use cooperative positioning where tags share information to improve collective accuracy~\cite{laadung22_novel_activ_passiv_two_way}. However, these approaches still rely on continuously powered infrastructure, preventing truly energy-neutral operation.

Further, Zhang et al.~\cite{zhang25_data_set_uwb_cooper_navig} explored \ac{UWB}-based cooperative localization in multi-agent systems, particularly for \ac{UAV} clusters, and published a dataset containing all distance measurements between the nodes. They demonstrated cooperative positioning approaches where multiple mobile nodes exploit ranging measurements both to fixed anchors and between themselves, achieving positioning accuracies below \qty{1}{\meter} in indoor environments and below \qty{2}{\meter} outdoors under various flight formations. Using \ac{DS-TWR}, their approach enables distance-based collaborative navigation without requiring external synchronization infrastructure, though at the cost of increased communication overhead when scaling to larger node counts. While this technique improves robustness, it still requires continuous node operation and does not address long-term energy sustainability.

In this work, we adopt the principle of concurrent transmissions from~\cite{corbalan20_ultra_wideb_concur_rangin}, but in a more relaxed fashion. Our approach allows sequential reception while requiring only a single transmitted message, as shown in \textsc{WakeLoc}~\cite{cortesi25_wakel}, thereby targeting both scalability and energy efficiency. Additionally, \textsc{WakeLoc}'s passive tags are used to opportunistically exploit the same messages for \ac{TDOA}~\cite{patru23_flext}. 

Comparing power consumption with localization quality, \ac{UWB} transceivers consume over \qty{100}{\milli\watt} during active listening~\cite{polonelli22_perfor_compar_decaw_dw100_dw300,istomin21_janus}, making battery-powered deployments challenging. Duty cycling reduces this issue by introducing latency and therefore limiting the responsiveness of the localizations~\cite{zhao21_uloc,luder25_anitr}. \textsc{WakeLoc}~\cite{cortesi25_wakel} showed that integrating wake-up radios with \ac{UWB} enables event-driven localization with similar accuracies as~\cite{patru23_flext,dotlic18_rangin_method_utiliz_carrier_frequen_offset_estim}, but did not address long-term energy neutrality.

\subsection{Wake-up Radio Technologies for Low-Power IoT}
\acp{WuR} have emerged as a key enabler for event-driven communication in energy-constrained \ac{IoT} systems~\cite{sutton15_zippy}. By maintaining ultra-low power listening capabilities while keeping the main radio in deep sleep, \acp{WuR} enables systems to achieve both responsiveness and energy efficiency.

Modern \ac{WuR} architectures achieve listening power consumption in the range of \qty{36}{\nano\watt} to \qty{100}{\micro\watt}~\cite{villani24_ultra_wideb_wake_up_receiv,polonelli21_ultra_low_power_wake_up}, orders of magnitude lower than conventional radio receivers. The sensitivity of \acp{WuR} typically ranges from \qty{-48}{\deci\belmilliwatt} to \qty{-75}{\deci\belmilliwatt}, with trade-offs between power consumption, reception range, and wake-duration. Recent implementations have demonstrated successful integration with various main radio technologies, including \ac{BLE}~\cite{piyare17_ultra_low_power_wake_up_radios}, and \ac{UWB}~\cite{schulthess25_wakem}.


A specialized \ac{WuR} \ac{IC} is the \textsc{FH101RF} from \textsc{Fraunhofer}, used in \textsc{WakeMod}~\cite{schulthess25_wakem}. It is the only commercially available \ac{WuR}, offering tri-band reception (\qty{433}{\mega\hertz}, \qty{868}{\mega\hertz} and \qty{2.4}{\giga\hertz}), idle listening consumption of \qty{6.9}{\micro\watt} and a sensitivity of \qty{-72.6}{\deci\belmilliwatt} whilst achieving wake-up latencies below \qty{55}{\milli\second} -- placing it close to state-of-the-art \acp{IC} from academia~\cite{villani24_ultra_wideb_wake_up_receiv, chen25_low_area_wake_up_receiv}, while keeping commercial availability and flexibility.

In our work, we exploit these advances by integrating \textsc{WakeMod} with its commercially available \ac{WuR} into a \ac{UWB} localization system, demonstrating that energy-neutral operation is feasible not only in theory but with practical, off-the-shelf components.

\subsection{Positioning Algorithms and Computational Trade-offs}
The choice of a three-dimensional positioning algorithm from known anchor positions has a significant impact on both localization accuracy and computational energy consumption, particularly for resource-constrained devices operating under strict energy budgets.

For multilateration problems, optimal solutions exist through eigenvalue decomposition approaches~\cite{larsson19_optim_trilat_is_eigen_probl,larsson25_singl_sourc_local_as_eigen_probl}. These methods achieve the statistically minimal error for given range measurements but require an eigendecomposition of a sparse matrix. In their comparison with state-of-the-art solvers~\cite{beck08_exact_approx_solut_sourc_local_probl,zhou10_closed_form_algor_least_squar_trilat_probl,luke17_simpl_global_conver_algor_nonsm}, Larsson et. al. showed the highest accuracy with \qty{34}{\centi\meter} in average over eight datasets, with the average solver having an error of \qty{70}{\centi\meter}. In terms of runtime complexity, it is \qty{47}{\times} more efficient than the average of the other solvers, and only \qty{16}{\percent} slower than the fastest solver~\cite{zhou10_closed_form_algor_least_squar_trilat_probl}.

Iterative methods, particularly the \ac{LM} algorithm~\cite{marquardt63_algor_least_squar_estim_nonlin_param,more78_leven_marquar}, offer a practical alternative with reduced computational requirements. While not guaranteed to find the optimum, \ac{LM}-iteration typically converges quickly for well-conditioned localization problems and can be implemented with limited memory usage. The trade-off between solution quality and computational cost makes iterative approaches interesting for energy-constrained systems.

In addition to the above methods, \ac{TDOA} positioning presents additional algorithmic challenges due to the hyperbolic nature of the constraint equations~\cite{spirito98_hyperbolic_posi,gust03positioning_tdoa}. Closed-form solutions exist for specific geometric configurations, but general cases require iterative or particle filter-based non-linear least squares solvers~\cite{gust03positioning_tdoa}.

We address these challenges by tailoring \ac{TDOA} and multilateration solvers to operate within the limited energy and computational resources of our platform, enabling scalable cooperative localization without sacrificing feasibility.

\section{System Architecture and Characterization}\label{sec:system-architecture}
\begin{figure}[htpb!]
  \centering
  \includesvg[width=0.8\columnwidth]{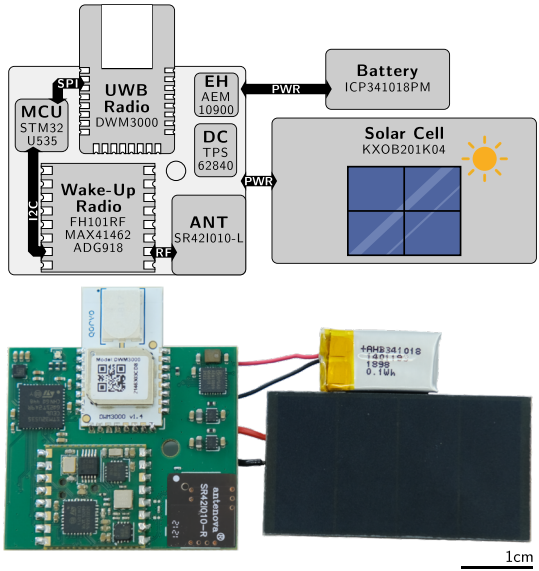}
  \vspace{-0.2cm}
  \caption{Overview of the developed sensor node used for anchors and tags.}\label{fig:hardware-overview}
  \vspace{-0.3cm}
\end{figure}

The proposed localization system described in this section consists of the anchor and tags, as shown in \cref{fig:hardware-overview}, a localization protocol based on \acp{WuR} and being processed on-line on the tags wanting to localize, and an energy-aware localization-scheduling algorithm.

\subsection{Hardware design}
The hardware architecture is based on energy harvesting from a solar cell, combined with a small \ac{LiPo} battery for buffering, ensuring autonomous operation. It combines an \ac{UWB} transceiver for localization with the \textsc{WakeMod} \ac{WuR} for efficient idle listening, all managed by a low-power \ac{MCU}.

The energy harvesting capability is based on an \textsc{e-peas} \textsc{AEM10900} boost converter with integrated \ac{MPPT} tracking. An \textsc{ANYSOLAR} \textsc{KXOB201K04TF} solar cell with a size of \qtyproduct{23x42}{\milli\meter} is used as energy source. The cell is selected with an open-loop voltage close to the harvesters' maximum input voltage, at which point it achieves its highest efficiency. 
A \qty{35}{\milli\ampere{}\hour} \ac{LiPo} battery is used to buffer energy during low-light conditions.
The \qty{1.8}{\volt} supply of the node itself is generated using the high-efficiency \qty{60}{\nano\ampere}-quiescent current buck converter \textsc{TPS62840} from \textsc{Texas Instruments}. 

The \textsc{STMicroelectronics} \textsc{STM32U535CEUx} microcontroller is based on a \textsc{ARM} \textsc{Cortex-M33} with \ac{FPU}, operates at frequencies up to \qty{160}{\mega\hertz}, and is equipped with \qty{512}{\kilo\byte} of flash memory and \qty{274}{\kilo\byte} of SRAM. It has a power consumption of \qty{16.3}{\micro\ampere\per\mega\hertz} and permits a low-power state of \qty{90}{\nano\ampere} in shutdown mode. 

Localization is carried out with the \textsc{Qorvo} \textsc{DWM3000} \ac{UWB} module, achieving a low power consumption around \qty{230}{\nano\ampere} in deep sleep. The configurations of the \ac{UWB} transceiver using~\cite{jiang24_demo} are optimized to achieve high range (channel \num{5}) while maintaining low TX consumption (\qty{128}{\bit} preamble). 



The final building block is the \ac{WuR} module \textsc{WakeMod}~\cite{schulthess25_wakem}. The open-source module is designed around the \textsc{Fraunhofer} \textsc{FH101RF} \ac{WuR} radio and the \textsc{Analog Devices} \textsc{MAX41462} \ac{ASK} transmitter, both connected through an \textsc{Analog Devices} \textsc{ADG918} \ac{RF} switch to the same antenna. The module designed for \qty{868}{\mega\hertz}, achieves a reception sensitivity of \qty{-72.6}{\deci\belmilliwatt} and a transmission power of up to \qty{2.78}{\deci\belmilliwatt} at \qty{1.8}{\volt}. With its consumption of \qty{6.9}{\micro\watt} when actively listening, the system remains in deep sleep and only wakes up upon receiving a \ac{WuC}.

\subsection{Localization Protocol and Algorithms}\label{subsec:accuracy-algorithms}
\begin{figure}[htpb!]
    \centering
    \includesvg[width=\columnwidth]{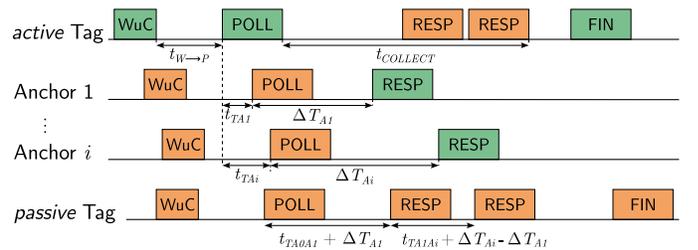}
    \vspace{-0.5cm}
    \caption{\textsc{WakeLoc} localization scheme as shown in~\cite{cortesi25_wakel}, combining \ac{TWR} with \ac{TDOA}. Messages in green are transmissions, orange represents receptions.}\label{fig:background-wakeloc}
    \vspace{-0.2cm}
\end{figure}

The employed localization method is based on \textsc{WakeLoc}~\cite{cortesi25_wakel}. Tags can obtain their position either by \textit{actively} initiating a localization procedure themselves (\textit{active} mode) or by \textit{passively} reusing the messages exchanged during another tag’s localization procedure (\textit{passive} mode) -- making it scalable in the amount of tags being part of the \ac{RTLS}. An overview of this scheme is illustrated in~\cref{fig:background-wakeloc}.

In \textit{active}-mode, a tag first wakes nearby anchors with a \ac{WuC} and then performs multilateration based on a nested \ac{CC-SS-TWR} exchange. Compared to naive approaches, this nested scheme reduces the number of required messages to \(N_A+1\) for a localization with \(N_A\) anchors. The tag computes its own position from the collected distance measurements and finally broadcasts its estimate.

The delay \(\Delta T_{Ai}\) of each anchor’s response depends on its index, using \(\Delta T_{Ai} = \Delta T_\textit{fix} +\Delta T \cdot\left[(i-1) \mod \hat{N}_A\right]\) to ensure scalability. The value of 
\(\hat{N}_A\) is chosen such that no two anchors within reception range reply simultaneously, requiring some spatial separation in their placement.

In \textit{passive}-mode, a tag reuses the localization exchange triggered by an \textit{active} tag. After receiving the initial \ac{WuC}, it listens for the poll and response messages of nearby anchors and applies \ac{TDOA} to derive its position. The missing information, namely the position of the \textit{active} tag, is obtained from the \textit{active} tag's final broadcast.

With this combination of \ac{TWR} and \ac{DL-TDOA}, \textsc{WakeLoc} achieves scalability for both the number of anchors and tags, whose low latency only depends on the \ac{WuR}, and allows for ultra-low power consumption on the anchors while they are actively listening.
However, the tag performing \textit{active} localization consumes more energy than in a standard \ac{TWR} system, as it additionally needs to transmit the \ac{WuC} and broadcast its position.

For localization, 
the tag acting as the \textit{active} one must execute its localization through multilateration, whilst a tag acting \textit{passively} applies \ac{TDOA}. Larsson et al.~\cite{larsson19_optim_trilat_is_eigen_probl} proved that the multilateration problem can be converted to an eigenvalue problem and thus be solved optimally (if the eigenvalues can be found exactly), achieving the statistically minimal error with the available measurements. Next to Larsson,  the \ac{LM}-iteration~\cite{marquardt63_algor_least_squar_estim_nonlin_param, more78_leven_marquar} was evaluated from the class of iterative (though non-optimal) solvers.

The \ac{TDOA} problem of the \textit{passive} tag, again a non-linear least squares problem, is solved as well using \ac{LM}-iteration. The different solvers implemented in this work are made available as an open-source library on \textsc{GitHub}\footnote{\url{https://github.com/ETH-PBL/WakeLoc-Algorithm}}.

\subsection{Energy-Aware AIMD Algorithm}\label{subsec:energyaware-algorithm}
The energy-aware and cooperative algorithm closely builds on \textsc{EcoTrack}~\cite{giordano23_energ_aware_adapt_sampl_self}. The algorithm is based on the principle of dynamically adapting the number of \textit{active} localizations according to the available harvested energy, with the battery state of charge serving as a proxy for energy income. The approach follows the \ac{AIMD} paradigm, well known from congestion control in computer networks. The algorithm increases the hourly localization rate linearly by one, as long as sufficient energy is available, but halves the rate (multiplicative decrease with factor \num{2}) once the battery indicates a lack of energy. The transitions between these states are controlled by a \ac{FSM} in combination with a metric function.

\subsubsection[FSM]{\Acf{FSM}}\label{subsubsec:energyaware-algorithm-fsm}
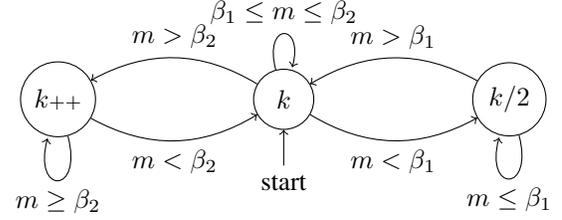
\begin{figure}[htpb!]
  \centering
  \begin{tikzpicture}
[align=center,node distance=3cm,
state/.style={circle, draw, minimum size=0.8cm}]
\node[state] (q1) {$k/2$};
\node[state, initial, initial where=below, left of=q1] (q2) {$k$};
\node[state, left of=q2] (q3) {$k{\scriptstyle++}$};
\draw [->] (q1) edge[loop below] node{$m\leq \beta_1$} (q1)
(q2) edge[loop above] node{$\beta_1\leq m\leq \beta_2$} (q2)
(q3) edge[loop below] node{$m\geq \beta_2$} (q3)
(q1) edge[bend right, above] node{$m>\beta_1$} (q2)
(q2) edge[bend right, below] node{$m<\beta_1$} (q1)
(q2) edge[bend right, above] node{$m>\beta_2$} (q3)
(q3) edge[bend right, below] node{$m<\beta_2$} (q2);
\end{tikzpicture}
  \caption{State diagram of the \acs{FSM}-based adaptive sampling algorithm: \(k/2\) means that \(k\) halves in the next step; \(k++\) means that \(k\) increases by a fixed rate in the next step, in our case by one; \(k\) means that \(k\) stays the same in the next step.}\label{fig:energyaware-algorithm-fsm}
\end{figure}

The \ac{FSM} of \textsc{EcoTrack}, illustrated in~\cref{fig:energyaware-algorithm-fsm}, operates on the variable \(k\), defined as the number of \textit{active} localizations per hour (of the tag running this instance of the \ac{FSM}): It either holds \(k\), increments \(k\) by one, or halves \(k\).

State transitions are triggered by evaluating \textsc{EcoTrack}'s metric function \(m\) against two thresholds \(\beta_1\) and \(\beta_2\), which determine whether the tag should increase or decrease the \textit{active} localization rate. In addition, a battery-level threshold \(\gamma\) ensures that the algorithm always increases \(k\) whenever the state of charge is high enough. Together, \(\beta_1, \beta_2\) and \(\gamma\) represent the tunable parameters of the \ac{FSM}.

\subsubsection{Metric Function}\label{subsubsec:energyaware-algorithm-metric}
The \textsc{EcoTrack}'s metric function \(m\) is defined in~\cref{eq:energyaware-algorithm-metric}. It evaluates the change in battery state \(b[t]\) between consecutive hours, normalized by the battery capacity \(B\). The metric is rewarded when the battery state is high or increasing, and penalized when the state is low or decreasing. In particular, when the battery charge remains above the boundary \(\gamma\), the algorithm always increases the localization rate, whereas a decrease in battery state triggers a multiplicative decrease of \(k\). This mechanism allows the system to adapt smoothly to fluctuating energy availability, maximizing the number of \textit{active} localizations while preserving long-term sustainability.

\begin{equation}\label{eq:energyaware-algorithm-metric}
    m = \underbrace{\vphantom{\begin{cases}0\\0\end{cases}} B\cdot(b[t] - b[t-1])}_{\text{batt. state difference}} - \underbrace{\vphantom{\begin{cases}0\\0\end{cases}} \left(\frac{1}{b[t]}-1\right)}_{\text{low batt. penalty}} + \underbrace{\begin{cases}
    \infty,& \text{if } b[t] \geq \gamma\\
    0,     & \text{otherwise}
\end{cases}}_{\text{high batt. reward}}
\end{equation}

A second version of this metric (\cref{eq:energyaware-algorithm-metric2}) keeps an upper bound on the number of \textit{active} localizations per day (system constraint, \textit{not} tuned), bounding the rate once this limit is reached.

\begin{equation}\label{eq:energyaware-algorithm-metric2}
    m_\textit{bounded} = \underbrace{\begin{cases}
    m,& \text{if } k < k_\textit{max}\\
    \min\left(m,\ \frac{\beta_1 + \beta_2}{2}\right), & \text{otherwise}
\end{cases}}_{\text{bound on max. localizations}}
\end{equation}

\subsubsection{Algorithm Tuning}\label{subsubsec:energyaware-algorithm-tuning}
The parameters \(\beta_1, \beta_2\) and \(\gamma\) are tuned off-line and globally to optimize performance subject to the constraints in~\cref{eq:energyaware-algorithm-constraints}. For each tag \(j\), the battery state at the end of the year must be at least \qty{10}{\percent}. During tuning, all batteries are initialized at \qty{10}{\percent} \ac{SoC} to model a conservative starting condition.
\begin{align}\label{eq:energyaware-algorithm-constraints}
  \begin{split}
    b_j[T] &\geq \qty{10}{\percent}\ \forall{j}\\
    b_j[\cdot] &\leq \qty{100}{\percent}\ \forall{j}\\
  \end{split}
\end{align}
The optimization problem in~\cref{eq:energyaware-algorithm-optimization} is formulated to maximize the mean number of \textit{active} and \textit{passive} localizations across all tags over one year. Let \(l_j[t] \in \mathbb{N}\) denote the number of localizations of tag \(j\) on day \(t\). Then,
\begin{equation}\label{eq:energyaware-algorithm-optimization}
  J = \max_{\{l_n[t]\}} \left(\frac{1}{N_T} \sum_{j=1}^{N_T} \sum_{t=1}^T l_j[t]\right) \quad \text{subj. to \cref{eq:energyaware-algorithm-constraints}}
\end{equation}
where \(N_T\) is the number of tags and \(T\) the number of days in the year.

Each tag then schedules every hour the \textit{active} localizations for the next hour in a uniformly at random manner, thereby improving temporal coverage. Furthermore, the algorithm regulates only \textit{active} localization decisions. In simulation, \textit{passive} localizations are always performed opportunistically, whenever energy is available, enabling tags to exploit ongoing communication exchanges and making the algorithm intrinsically cooperative.
\section{Experimental Setup}\label{sec:exp-setup}
To validate the system's performance and accuracy, several evaluations were conducted targeting four core aspects: localization accuracy across different solvers, performance and power consumption of the ultra-low-power sensor node, and system behavior in both controlled and real-world deployments, including quadruped-robot evaluation and large-scale simulation. The following subsections describe each experimental setup in detail.

\subsection{Localization Accuracy}\label{subsec:accuracy-setup}
In order to evaluate the different algorithmic solvers for localization (Larsson and \ac{LM}-iteration for multilateration and \ac{LM}-iteration for \ac{TDOA} as described in \cref{subsec:accuracy-algorithms}), an experiment was set up in a room measuring \qtyproduct{8.6x7.6}{\meter}. A \textsc{Bosch} \textsc{GLM150-27C} laser distance meter and a \textsc{Vicon} motion capture system were employed to provide the ground truth.

\begin{figure}[htpb!]
  \centering
  \includesvg[width=0.8\columnwidth]{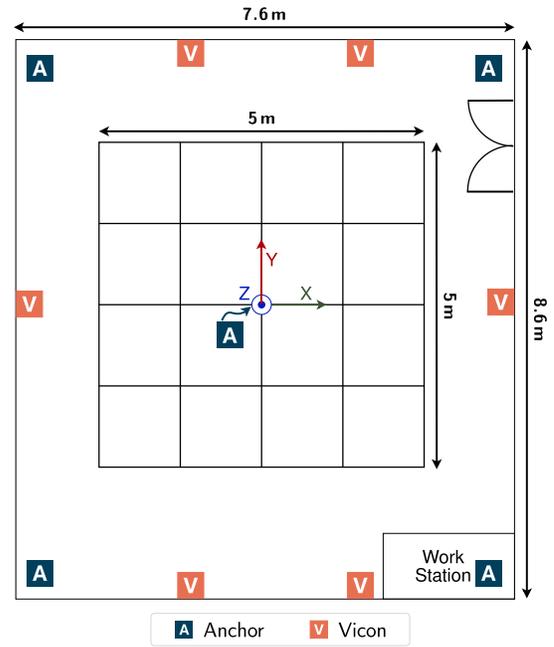}
  \caption{Experimental setup for evaluating the localization accuracy. \(A\) represents anchors, \(V\) the cameras of the \textsc{Vicon} system.}\label{fig:accuracy-setup}
  \vspace*{-0.2cm}
\end{figure}

To minimize the \ac{GDOP}, five anchors were strategically positioned on the given infrastructure to attain optimal spatial coverage~\cite{ding25_optim_placem_minim_gdop_with,yang22_high_precis_uwb_based_local} (\ac{GDOP} of \num{0.090}). Two tags, one \textit{active} and one \textit{passive}, were placed at varying heights within a \qtyproduct{5x5}{\meter} grid. A total of \num{200} distinct positional combinations were tested. For this, first the \textit{active} tag was moved on the grid with the \textit{passive} fixed in position, and then vice versa. Around \num{480} distinct measurements were conducted for each position combination, yielding a dataset of \num{95457} measurements.  A first preliminary measurement of each anchor was employed to calibrate its antenna delay. The measurement setup is shown in~\cref{fig:accuracy-setup}

\begin{table*}[!b]
    \vspace*{-0.5cm}
    \renewcommand{\arraystretch}{1.2}
    \centering
    \caption{Accuracy results depending on the localization type and used algorithm. Values are given in \unit{\centi\meter}.}\label{tab:accuracy-results}
    \begin{tabular}{@{}lrrcrrr@{}}
    \toprule
     & \multicolumn{2}{c}{\textsc{WakeLoc} \textit{active} (\acs{TWR})} & \phantom{a} & \multicolumn{3}{c}{\textsc{WakeLoc} \textit{passive} (\acs{TDOA})}\\
     \cmidrule{2-3} \cmidrule{5-7}
     & Larsson & \acs{LM} & & \acs{TDOA} with \acs{GT} & \acs{TDOA} with Larsson & \acs{TDOA} with \acs{LM} \\
    \midrule
    avg & \qty{21.68}{\centi\meter} & \qty{21.89}{\centi\meter} & & \qty{33.05}{\centi\meter} & \qty{25.66}{\centi\meter} & \qty{25.70}{\centi\meter} \\
    md & \qty{16.13}{\centi\meter} & \qty{16.13}{\centi\meter} & & \qty{27.49}{\centi\meter} & \qty{19.81}{\centi\meter} & \qty{19.86}{\centi\meter} \\
    $\sigma(\cdot)$ & \qty{17.02}{\centi\meter} & \qty{17.13}{\centi\meter} & & \qty{24.84}{\centi\meter} & \qty{18.97}{\centi\meter} & \qty{19.05}{\centi\meter} \\
    $\min{\|\cdot\|}$ & \qty{0.28}{\centi\meter} & \qty{0.27}{\centi\meter} & & \qty{0.35}{\centi\meter} & \qty{0.42}{\centi\meter} & \qty{0.26}{\centi\meter} \\
    $\max{\|\cdot\|}$ & \qty{497.72}{\centi\meter} & \qty{345.75}{\centi\meter} & & \qty{485.79}{\centi\meter} & \qty{464.38}{\centi\meter} & \qty{469.36}{\centi\meter} \\
    \bottomrule
    \end{tabular}
\end{table*}

In a second evaluation, the \ac{RTLS} has been evaluated in a realistic scenario as depicted in~\cref{fig:robodog-results}. The experimental setup included an office with a size of \qtyproduct{5.7x7.0}{\meter} and an adjacent corridor of size \qtyproduct{3.8x14.7}{\meter}, with a total space of \qty{91}{\meter\squared}. In this real-world use case, the tags were mounted on the top of a \textsc{Unitree} \textsc{A1} quadruped robot equipped with the sensor backpack described in~\cite{plozza24_auton_navig_dynam_human_envir}. An \textit{active} tag and a \textit{passive} tag were positioned at a height of \qty{53.4}{\centi\meter} from the ground and separated by \qty{20.7}{\centi\meter}. The experiment involved the robot leaving the office and traversing the corridor. A total of nine anchors and two tags were deployed in this setup. The robot’s \textsc{Hokuyo} \textsc{UTM-30LX-EW} 2D LiDAR with \ac{AMCL} as described in~\cite{plozza24_auton_navig_dynam_human_envir} was used as ground truth, as its centimeter-level accuracy exceeds the precision required to benchmark the \ac{RTLS}.

\subsection{Computational Cost}\label{subsec:compute-setup}
The recorded dataset collected as described in~\cref{subsec:accuracy-setup} was used to evaluate the three microcontroller implementations (optimal/iterative multilateration, iterative \ac{TDOA}) on the \ac{MCU}. In addition, four distinct solver tolerances (\num{e-2}, \num{e-4}, \num{e-6}, and \num{e-8}) were used during evaluation. On the \ac{LM} algorithms, the tolerance is used as the relative tolerance for early stopping the iteration, whereas on the Larsson algorithm, it is used inside the inverse iteration during eigenvector calculation. This approach allows for a quantitative examination of both localization statistical accuracy (mean error, variance) and algorithmic performance (computational time, resource demands) - both evaluated on the developed sensor node. The computation time was measured using the \ac{DWT} peripheral of the \textsc{ARM} \textsc{Cortex-M33} \ac{MCU}.

Three scenarios were evaluated for the \ac{LM}-\ac{TDOA} solver:
\begin{enumerate*}[label=\textit{\roman*)}]
\item Assuming the known position of the \textit{active} tag (classical TDOA),
\item using the position of the \textit{active} tag estimated by the Larsson multilateration, and
\item using \ac{LM} multilateration.
\end{enumerate*}

\subsection{Power Consumption}\label{subsec:powerconsumption-setup}
The power consumption of the nodes was analyzed using a \textsc{Keysight} \textsc{N6705C} DC Power analyzer fitted with three \textsc{N6781A} source measurement units. Three nodes have been attached to the power analyzer simultaneously on three different channels, one configured as an anchor, the others as \textit{active} and \textit{passive} tags respectively.
Using a sampling frequency of \qty{48.8}{\kilo\hertz}, power profiles of the three nodes were recorded over a \qty{30}{\second} window and multiple localizations.

\subsection{Energy-Aware Simulation-Based Evaluation}\label{subsec:simulation-setup}
Next to the real-world deployment, we created a simulation framework to estimate the behavior of a large-scale deployment with respect to the proposed energy-aware localization algorithm. 

Geometrically, the simulation is based on the \textsc{Dept} environment of the \textsc{Cloves} testbed~\cite{molteni22_cloves}, covering an area of \qty{6382}{\meter\squared} consisting of two larger open spaces connected by a narrow corridor. The deployment includes \num{89} anchor positions. In addition, scenarios with \num{5}, \num{20}, and \num{100} randomly placed tags were simulated over one year. For a successful localization, a minimum of five successful anchor responses was assumed.

\begin{figure}[htpb!]
  \centering
  \stackinset{l}{10pt}{b}{115pt}{(a)}{%
    \stackinset{l}{10pt}{b}{5pt}{(b)}{%
      \includesvg[width=0.8\columnwidth]{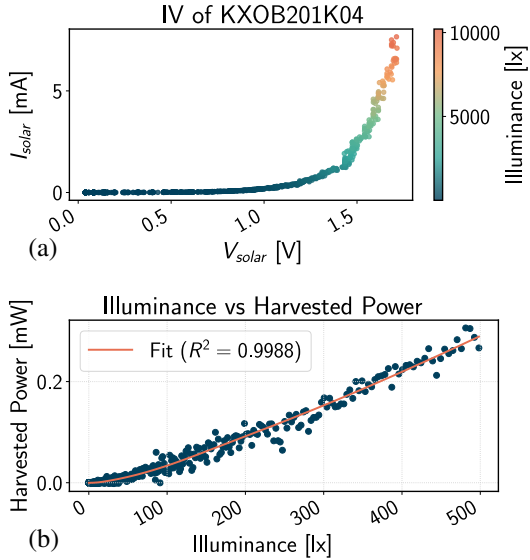}%
    }%
  }%
  \vspace*{-0.3cm}
  \caption{Solar cell and harvester characteristics. (a) shows the IV-curve of the \textsc{KXOB201K04} solar cell, and (b) shows the measured relation between light intensity and harvested power into a \qty{3.7}{\volt} sink together with a fitted polynomial function.}\label{fig:simulation-harvester}
  \vspace*{-0.2cm}
\end{figure}

\begin{figure*}[!t]
  \centering
  \stackinset{l}{20pt}{b}{23pt}{(a)}{%
    \stackinset{l}{275pt}{b}{23pt}{(b)}{%
      \includesvg[width=\textwidth]{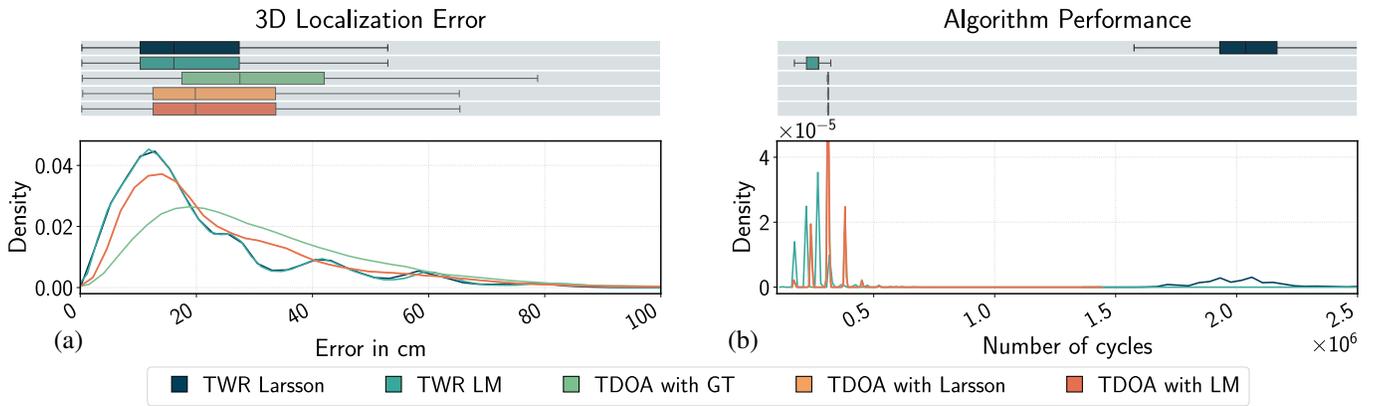}%
    }%
  }%
  \vspace*{-0.7cm}
  \caption{Localization accuracy (a) and computational cost (b) of different multilateration and \ac{TDOA} solvers.}
  \label{fig:localization}
  \vspace*{-0.5cm}
\end{figure*}

As no long-term dataset was available for our specific harvester and solar cell, the indoor solar harvesting dataset presented in~\cite{sigrist19_datas} was used, providing over two years of indoor solar harvesting measurements from six different locations inside an office building. At each location, the illuminance was recorded at a frequency of \qty{1}{\hertz}. For the purpose of this evaluation, the complete year 2018 was considered with its five monitored positions. These locations represent a wide range of lighting conditions, from window-facing offices to hallways with very little natural light.

To adapt the dataset to our sensor node electronics, first, the \textsc{AEM10900} together with the \textsc{KXOB201K04} solar panel was characterized in a solar testbed connected to two \textsc{Keysight} \textsc{B2902A} source-measurement units. The relation was measured across \num{1000} illuminance levels from \qty{0.2}{\lux} up to \qty{10}{\kilo\lux} at a \ac{MPPT} of \qty{75}{\percent}, as shown in~\cref{fig:simulation-harvester}.

Three sigmoid-blended polynomials were fitted to the measured data for the regions \qty{0.2}{\lux}–\qty{15}{\lux}, \qty{15}{\lux}–\qty{1500}{\lux}, and \qty{1500}{\lux}–\qty{10}{\kilo\lux}. The resulting illuminance-to-harvested-power model achieved an $R^2$ of \num{0.9988}. Using this model, the illuminance measurements from Sigrist et al.~\cite{sigrist19_datas} datasets were transformed into the power harvestable by our system and downsampled to one sample per minute. 
The daily harvestable energy ranges from only \(\num{0.31}\pm\qty{0.13}{\joule}\) in dim environments to up to \(\num{18.51}\pm\qty{20.65}{\joule}\) in bright settings, with typical indoor conditions giving about \(\num{1.81}\pm\qty{1.21}{\joule}\) per day.


Based on these considerations, the simulation was set up as follows: each anchor and tag was equipped with an empty \qty{35}{\milli\ampere{}\hour} battery in the simulation (charged to \qty{0}{\percent}), as well as a randomly assigned energy harvesting profile from the dataset. Battery self-discharge was modeled according to the \textsc{renata} userguide~\cite{renata_userguide}: \qty{4}{\micro\ampere} leakage at full capacity and \qty{1}{\micro\ampere} at \qty{30}{\percent} \ac{SoC}, interpolated linearly in between.

\begin{table*}[!b]
    \vspace*{-0.5cm}
    \centering
    \renewcommand{\arraystretch}{1.2}
    \caption{Real-world evaluation of the accuracy against the \ac{LiDAR} of the quadruped robot. Values are given in \unit{\meter}.}\label{tab:robodog-results}
    \begin{tabular}{@{}lrrrrcrrrr@{}}
     \toprule
     & \multicolumn{4}{c}{\textsc{WakeLoc} \textit{active} (\acs{TWR})} & \phantom{a} & \multicolumn{4}{c}{\textsc{WakeLoc} \textit{passive} (\acs{TDOA})}\\
     \cmidrule{2-5} \cmidrule{7-10}
     & \multicolumn{2}{c}{Larsson} & \multicolumn{2}{c}{\acs{LM}} & & \multicolumn{2}{c}{\acs{TDOA} with Larsson} & \multicolumn{2}{c}{\acs{TDOA} with \acs{LM}} \\
     & w outliers & w/o outliers & w outliers & w/o outliers & & w outliers & w/o outliers & w outliers & w/o outliers\\
     \midrule
     avg               & \qty{0.47}{\meter} & \qty{0.43}{\meter} & \qty{0.48}{\meter} & \qty{0.43}{\meter} & & \qty{6.81}{\meter} & \qty{4.03}{\meter} & \qty{5.76}{\meter} & \qty{6.91}{\meter} \\
     md                & \qty{0.34}{\meter} & \qty{0.37}{\meter} & \qty{0.36}{\meter} & \qty{0.35}{\meter} & & \qty{0.46}{\meter} & \qty{0.43}{\meter} & \qty{0.52}{\meter} & \qty{0.46}{\meter} \\
     $\sigma(\cdot)$   & \qty{0.49}{\meter} & \qty{0.25}{\meter} & \qty{0.35}{\meter} & \qty{0.24}{\meter} & & \qty{16.64}{\meter} & \qty{10.59}{\meter} & \qty{15.18}{\meter} & \qty{16.83}{\meter} \\
     $\min{\|\cdot\|}$ & \qty{0.05}{\meter} & \qty{0.08}{\meter} & \qty{0.05}{\meter} & \qty{0.06}{\meter} & & \qty{0.04}{\meter} & \qty{0.04}{\meter} & \qty{0.03}{\meter} & \qty{0.03}{\meter} \\
     $\max{\|\cdot\|}$ & \qty{4.07}{\meter} & \qty{1.31}{\meter} & \qty{2.02}{\meter} & \qty{1.27}{\meter} & & \qty{107.33}{\meter} & \qty{64.74}{\meter} & \qty{68.81}{\meter} & \qty{68.81}{\meter} \\
     \bottomrule
    \end{tabular}
\end{table*}

At every simulation step (one minute), first, every node received harvested energy corresponding to its harvesting profile.
Then, tags were processed in randomized order to ensure fairness for their energy-aware algorithm. For each tag, the decision to initiate \textit{active} localization was made using the algorithms introduced in~\cref{subsec:energyaware-algorithm}. Both strategies were compared to a constant-rate algorithm (with the rate being a tunable parameter), where each tag tries to perform a static amount of \textit{active} localizations per hour.
For both the \ac{AIMD}-based and constant-rate algorithms, the parameters were tuned only at the average-light positions P06 and P13 prior to the full-year simulations and in a simulation with \num{100} randomly and uniformly distributed tags in the simulation environment. The limit for the constant-rate algorithm to fulfill the constraints in~\cref{eq:energyaware-algorithm-constraints} is one \textit{active} localization per hour. The bound for the bounded-\ac{AIMD} is set to \num{6} localizations per hour.


If a tag decides to perform an \textit{active} localization, anchors within \qty{20}{\meter} \ac{LOS} or \qty{5}{\meter} \ac{NLOS} will be considered. The number of anchors within reach of each other \(\hat{N}_A\) has been set to \num{10}. Each anchor's energy is reduced according to the power characterization presented in~\cref{subsec:powerconsumption-result}. A response is considered successful only if the anchor's battery remained non-empty afterwards. If at least five anchors respond successfully, the localization is considered successful; otherwise, the attempt is considered a failure.
Neighboring tags within the range of the same anchors are able to perform \textit{passive} localization in the same timestep. If more than five successful anchor responses are observed, the \textit{passive} localization succeeds.
Finally, the tag's battery state was adjusted accordingly to the consumption.
\begin{figure*}[!t]
  \centering
  \stackinset{l}{20pt}{b}{23pt}{(a)}{%
    \stackinset{l}{275pt}{b}{23pt}{(b)}{%
      \includesvg[width=\textwidth]{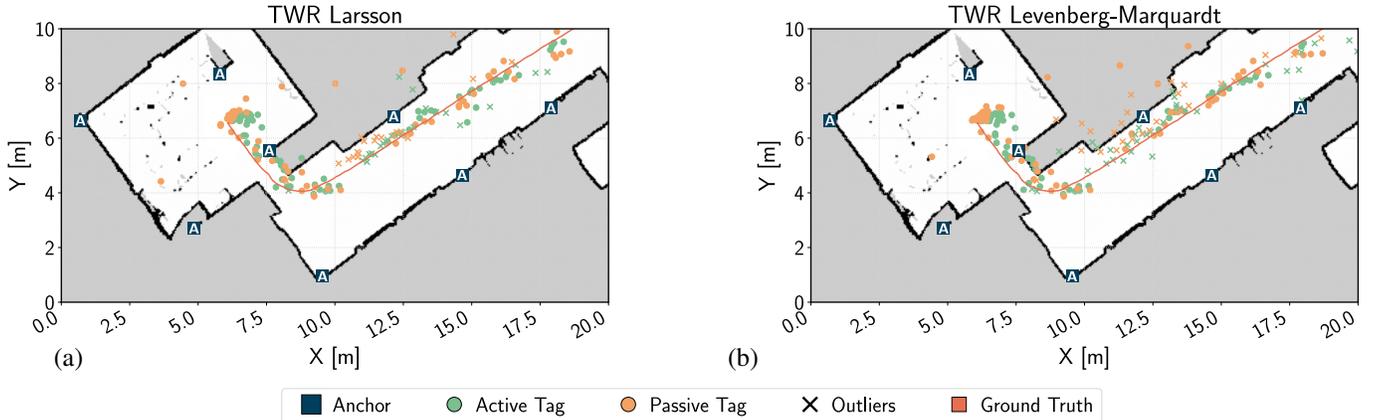}%
    }%
  }%
  \vspace*{-0.7cm}
  \caption{Localization of a dynamic object in a real-world environment. (a) shows the results when using Larsson's multilateration method, (b) when using \ac{LM}-iteration. Outliers are marked where localization failed due to having very few anchors (three or fewer) or a lack of convergence within \qty{20} iterations.}\label{fig:robodog-results}
  \vspace*{-0.5cm}
\end{figure*}

\section{Results}\label{sec:results}
Using the results obtained through the measurements as described in~\cref{sec:exp-setup}, this section presents their results and discusses their findings. The evaluations contains both controlled and real-world deployments, providing comprehensive validation of the system's capabilities. The following subsections detail the results for localization accuracy, computational performance, power consumption characteristics, and large-scale energy-aware and cooperative behavior.

\subsection{Localization Accuracy}\label{subsec:accuracy-result}
The resulting position results obtained using the measurements described in~\cref{subsec:accuracy-setup} were almost identical across all tolerances (sub-\unit{mm} changes) and
%
are summarized in \cref{tab:accuracy-results} for a solver tolerance of \num{e-2} and shown in \cref{fig:localization}. It can be seen that the maximum error for the multilateration with the \ac{LM}-iteration is lower, while the average error of \qty{21.89}{\centi\meter} is slightly higher than with the Larsson algorithm (\qty{21.68}{\centi\meter} on average, median \qty{16.13}{\centi\meter}). For the \ac{TDOA} solver, utilizing the active tag's position estimated via multilateration yields higher localization accuracy than relying on the initiator's ground truth. While classic \ac{TDOA} based on the ground truth resulted in an average error of \qty{33.05}{\centi\meter}, employing multilateration estimates from the Larsson algorithm and the \ac{LM}-iteration improved the accuracy to \qty{25.66}{\centi\meter} and \qty{25.70}{\centi\meter}, respectively. This phenomenon aligns with observations made in~\cite{cortesi25_wakel}, which attribute such performance gains to the availability of additional measurement data.

These results highlight two important points: First, the Larsson algorithm performs marginally better on average than the \ac{LM}-iteration, consistent with its theoretical property of solving for the optimum. In rare cases, however, the \ac{LM}-iteration achieves better convergence, as evidenced by its smaller maximum error. The large maximum error for the Larsson algorithm occurred when it failed to converge within 1000 power method iterations (\qty{427}{\milli\second}), unable to properly solve for the maximum eigenvalue. This convergence failure can occur when iteratively solving for the eigenvalues in poorly conditioned scenarios, particularly when sensor geometry is unfavorable or when significant measurement noise is present~\cite{malivert23_compar_improv_multil_solvin_simul, frisch25_why_not_use_tdoa}. Second, there is a negligible difference in the \ac{TDOA} results when using either multilateration method, as also seen in~\cref{fig:localization} where the curves for "\ac{TDOA} with \ac{LM}" and "\ac{TDOA} with Larsson" almost perfectly overlap.

\begin{table*}[!b]
    \vspace*{-0.5cm}
    \renewcommand{\arraystretch}{1.2}
    \centering
    \caption{Computational evaluation depending on the localization type and used algorithm. Values are given in cycles and \unit{\milli\second}. For readability, the table lists execution time in ms, derived from the \qty{144}{\mega\hertz} \ac{MCU} clock.}\label{tab:accuracy-computation}
    \begin{tabular}{@{}l rr rr c rr rr rr@{}}
    \toprule
     & \multicolumn{4}{c}{\textsc{WakeLoc} \textit{active} (\acs{TWR})} & \phantom{a} & \multicolumn{6}{c}{\textsc{WakeLoc} \textit{passive} (\acs{TDOA})} \\
     \cmidrule{2-5} \cmidrule{7-12}
     & \multicolumn{2}{c}{Larsson} & \multicolumn{2}{c}{\acs{LM}} & & 
       \multicolumn{2}{c}{\acs{TDOA} with \acs{GT}} & \multicolumn{2}{c}{\acs{TDOA} with Larsson} & \multicolumn{2}{c}{\acs{TDOA} with \acs{LM}} \\
    \midrule
    avg          & \num{2156437} & \qty{14.98}{\milli\second} & \num{267244} & \qty{1.86}{\milli\second} & & \num{317035} & \qty{2.20}{\milli\second} & \num{312427} & \qty{2.17}{\milli\second} & \num{312445} & \qty{2.17}{\milli\second} \\
    md           & \num{2037673} & \qty{14.15}{\milli\second} & \num{271548} & \qty{1.89}{\milli\second} & & \num{311197} & \qty{2.16}{\milli\second} & \num{311630} & \qty{2.16}{\milli\second} & \num{311630} & \qty{2.16}{\milli\second} \\
    $\sigma(\cdot)$ &   \num{777012} & \qty{5.40}{\milli\second} & \num{67046} & \qty{0.47}{\milli\second} & & \num{70249} & \qty{0.49}{\milli\second} & \num{54194} & \qty{0.38}{\milli\second} & \num{54194} & \qty{0.38}{\milli\second} \\
    $\min{\|\cdot\|}$ & \num{1435471} & \qty{9.97}{\milli\second} & \num{122750} & \qty{0.85}{\milli\second} & & \num{170021} & \qty{1.18}{\milli\second} & \num{170902} & \qty{1.19}{\milli\second} & \num{170900} & \qty{1.19}{\milli\second} \\
    $\max{\|\cdot\|}$ & \num{61419270} & \qty{426.52}{\milli\second} & \num{2495594} & \qty{17.33}{\milli\second} & & \num{1434213} & \qty{9.96}{\milli\second} & \num{1432733} & \qty{9.95}{\milli\second} & \num{1433041} & \qty{9.95}{\milli\second} \\
    \bottomrule
    \end{tabular}
\end{table*}

The evaluation in a real-world scenario on the quadruped robot (see~\cref{fig:robodog-results} and~\cref{tab:robodog-results}) shows slightly poorer results. On the one hand, the anchors are not optimally placed in terms of their \ac{GDOP}. On the other hand, anchors for which there is no direct \ac{LOS} are also taken into account in the position determination -- even though these could in a further step be excluded by using the \ac{UWB} messages' \ac{CIR}. The crossed-out position estimates are recognized by the localization algorithm as either due to a lack of convergence or an insufficient number of available anchors (\(<\num{4}\)) and classified as outliers.

The results of the \textit{active} tags continue to show that the Larsson algorithm and the \ac{LM}-iteration provide comparable accuracy, with average deviations of \qty{43}{\centi\meter} (both without taking into account the estimates identified as outliers). Notably, the \ac{LM}-iteration has a lower variance due to occasional convergence failures of the Larsson algorithm, resulting in smaller maximum errors (\qty{2.02}{\meter} versus \qty{4.07}{\meter}). The results of the \textit{passive} tags suffer significantly from the less accurate position determinations of the \textit{active} tags and from the \ac{NLOS} conditions -- an effect that has already been demonstrated in~\cite{zandian18_nlos_detec_mitig_differ_local}. While the median accuracy reaches up to \qty{0.46}{\meter}, the average and, in particular, the maximum errors are significantly higher. With knowledge of the building plan, these errors could be detected and reduced.

\subsection{Computational Cost and Performance}\label{subsec:compute-result}
The results in this section are based on the measurements described in~\cref{subsec:compute-setup}. For all solver tolerances, the average number of \ac{MCU} cycles varied by about \qty{15}{\percent}, between a tolerance of \num{e-2} and one of \num{e-8}. It is therefore notable that convergence with the iterative solvers -- if it succeeds -- happens quickly, which can be observed due to the minimal changes in the estimation errors across different solver tolerances. Here, it is important to distinguish between the number of algorithm iterations required for convergence and the resulting number of cycles. The average number of iterations (for both multilateration, as well as \ac{TDOA}) the \acl{LM} solver took to converge is four iterations. A cut-off at \num{20} iterations is used to identify non-converging cases. When multilaterating, non-convergence occurred in \num{8} out of \num{95457} cases for both the \ac{LM}-iteration and the Larsson algorithm. In \ac{TDOA}, non-convergence occurred in \num{96} cases when using ground truth data of the \textit{active} tag, \num{97} when using the estimate, respectively. 

The computational cost of the different algorithms is listed in \cref{tab:accuracy-computation} and \cref{fig:localization}b). The Larsson algorithm requires, on average, \num{2156437} clock cycles. The \ac{LM}-based multilateration takes \num{267244} cycles. For \ac{TDOA}, the \ac{LM}-based solver of the \textit{passive} tag takes about \num{312427} cycles on average. In~\cref{fig:localization}b), one can clearly see the individual iterations of the iteration-based solvers.

\subsection{Power Consumption}\label{subsec:powerconsumption-result}
The power consumption of the anchors and tags (one acting as an \textit{active} tag, one as a \textit{passive} tag) in deep sleep, with the \ac{WuR} actively listening, is \qty{7.84}{\micro\watt} -- \qty{1.15}{\micro\watt} higher than the consumption of the \textsc{WakeMod} alone.

\begin{figure*}[!t]
  \centering
  \stackinset{l}{10pt}{b}{20pt}{(a)}{%
    \stackinset{l}{185pt}{b}{20pt}{(b)}{%
      \stackinset{l}{350pt}{b}{20pt}{(c)}{%
        \includesvg[width=\textwidth]{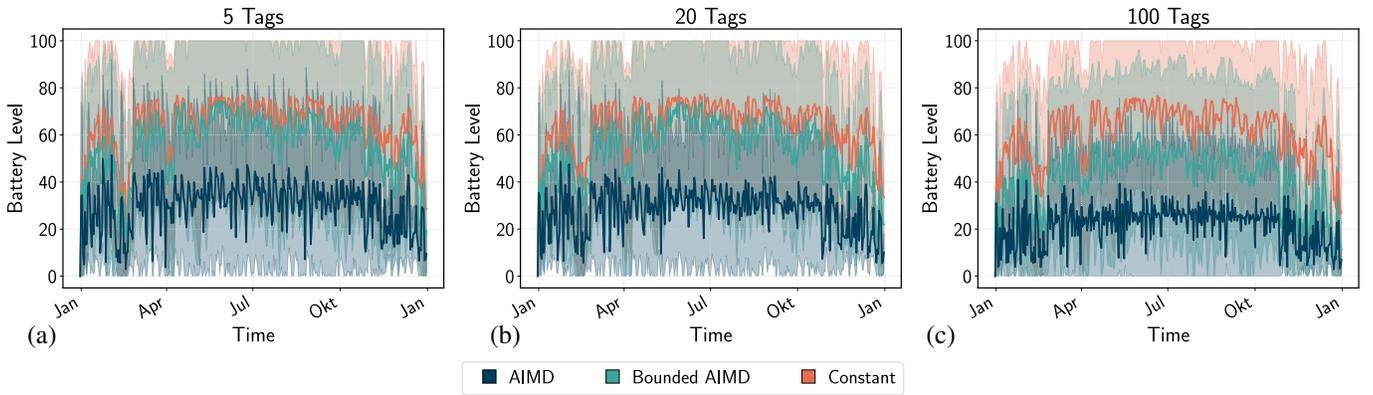}%
      }%
    }%
  }%
  \vspace*{-0.7cm}
  \caption{Average tag battery \ac{SoC} over the simulated year for simulations with (a) \num{5}, (b) \num{20}, or (c) \num{100} tags. Shaded area shows the standard deviation.}\label{fig:energy-aware-battery-results}
  \vspace*{-0.5cm}
\end{figure*}

\begin{figure*}[!b]
  \vspace*{-0.5cm}
  \centering
  \stackinset{l}{10pt}{b}{20pt}{(a)}{%
    \stackinset{l}{185pt}{b}{20pt}{(b)}{%
      \stackinset{l}{350pt}{b}{20pt}{(c)}{%
        \includesvg[width=\textwidth]{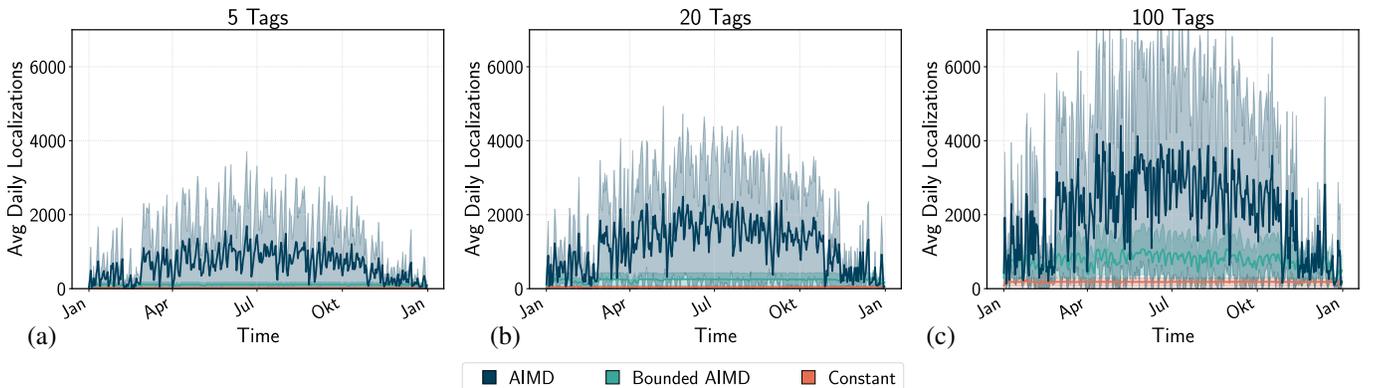}%
      }%
    }%
  }%
  \vspace*{-0.7cm}
  \caption{Average localizations per day and tag for simulations with (a) \num{5}, (b) \num{20}, or (c) \num{100} tags. Shaded area shows the standard deviation.}\label{fig:energy-aware-localization-results}
\end{figure*}

\cref{tab:powermeasurement} shows a breakdown of the energy consumption of the sensor nodes during a localization cycle, broken down by device type and localization method. \textit{Active} tags consume the most energy at \qty{3.22}{\milli\joule} per localization, as they actively transmit the \ac{WuC} and initiate measurements. \textit{Passive} tags require significantly less energy (\qty{951.16}{\micro\joule}), as they only receive and process signals from \textit{active} tags. The consumption of the anchors depends on the anchors' node number: Each localization causes \qty{338.30}{\micro\joule} of basic energy plus \qty{15.28}{\micro\joule} per localization, bound by \(\hat{N}_A\) as described in~\cref{subsec:accuracy-algorithms}.

The duration of the localization dictates the latency as well as the maximal localization rate. With a duration and latency of \qty{84.52}{\milli\second}, the maximal localization frequency is \qty{11.8}{\hertz}.

\begin{table}[!htpb]
  \vspace*{-0.3cm}
  \renewcommand{\arraystretch}{1.2}
  \centering
  \caption{Breakdown of the sensor nodes' power consumption.}\label{tab:powermeasurement}
  \begin{tabular}{@{}llrrr@{}}
    \toprule
                                  &          & \textit{active} Tag & \textit{passive} Tag & Anchor \\ 
    \midrule
    \multirow{4}{*}{Local.} & \multirow{2}{*}{Energy}   & \multirow{2}{*}{\qty{3.22}{\milli\joule}} & \multirow{2}{*}{\qty{951.16}{\micro\joule}} & \(\qty{338.30}{\micro\joule}+\qty{15.28}{\micro\joule}\) \\ &&&&\(\cdot\left[(i-1) \mod \hat{N}_A\right]\)\\
                                  & \multirow{2}{*}{Duration} & \multirow{2}{*}{\qty{84.52}{\milli\second}}  & \multirow{2}{*}{\qty{34.18}{\milli\second}}  & \multicolumn{1}{l@{}}{\(\qty{30.55}{\milli\second}+\qty{290}{\micro\second}\)} \\ &&&&\(\cdot\left[(i-1) \mod \hat{N}_A\right]\)\\
    \midrule
    Sleep                         & Power    & \qty{7.84}{\micro\watt} & \qty{7.84}{\micro\watt} & \qty{7.84}{\micro\watt}\\
    \bottomrule
  \end{tabular}
  \vspace*{-0.5cm}
\end{table}

\begin{table*}[!t]
    \renewcommand{\arraystretch}{1.2}
    \centering
    \caption{Statistics of the localizations per day of tags and anchors.}\label{tab:energy-aware-localization-results}
    \begin{tabular}{@{}lrrrcrrrcrrr@{}}
     \toprule
     & \multicolumn{3}{c}{\acs{AIMD}} & \phantom{a} & \multicolumn{3}{c}{Bounded \acs{AIMD}} & \phantom{a} & \multicolumn{3}{c}{Constant Rate}\\
     \cmidrule{2-4} \cmidrule{6-8} \cmidrule{10-12}
     \# Tags& \multicolumn{1}{r}{5} & \multicolumn{1}{r}{20} & \multicolumn{1}{r}{100} && \multicolumn{1}{r}{5} & \multicolumn{1}{r}{20} & \multicolumn{1}{r}{100} && \multicolumn{1}{r}{5} & \multicolumn{1}{r}{20} & \multicolumn{1}{r}{100} \\
     & T, A & T, A & T, A & & T, A & T, A & T, A & & T, A & T, A & T, A\\
     \midrule
avg               & \num{634}, \num{199} & \num{1173}, \num{1056} & \num{2031}, \num{2554}  && \num{105}, \num{35} & \num{234}, \num{186} & \num{761}, \num{811}  && \num{19}, \num{7} & \num{42}, \num{34} & \num{180}, \num{182}  \\ 
md                & \num{154}, \num{0} & \num{569}, \num{274} & \num{1092}, \num{1814}  && \num{141}, \num{0} & \num{281}, \num{142} & \num{805}, \num{848}  && \num{24}, \num{0} & \num{48}, \num{24} & \num{193}, \num{184}  \\ 
$\sigma(\cdot)$   & \num{886}, \num{568} & \num{1375}, \num{1397} & \num{2398}, \num{2564}  && \num{60}, \num{74} & \num{167}, \num{168} & \num{550}, \num{568}  && \num{10}, \num{26} & \num{29}, \num{38} & \num{123}, \num{140}  \\ 
min               & \num{0}, \num{0} & \num{0}, \num{0} & \num{0}, \num{0}  && \num{0}, \num{0} & \num{0}, \num{0} & \num{0}, \num{0}  && \num{0}, \num{0} & \num{0}, \num{0} & \num{0}, \num{0}  \\ 
max               & \num{4211}, \num{4232} & \num{6194}, \num{7935} & \num{14354}, \num{14341}  && \num{295}, \num{1681} & \num{576}, \num{1739} & \num{3108}, \num{5899}  && \num{67}, \num{1281} & \num{96}, \num{1250} & \num{2807}, \num{3083}  \\ 
     \bottomrule
    \end{tabular}
    \vspace*{-0.5cm}
\end{table*}

\subsection{Energy-Aware Simulation}\label{subsec:simulation-result}
The evaluation of the simulation focused on the achievable total number of localizations per tag and day, as well as the \ac{SoC} of the battery over the year.

The results in \cref{fig:energy-aware-battery-results} show the average \ac{SoC} of the tags over the entire simulation period. Regardless of the algorithm selected and the network size, the tags achieved to maintain an average charge of \qty{7}{\percent} at the end of the simulated year. The \ac{AIMD} consumed most energy, followed by the bounded-\ac{AIMD} and the constant rate algorithm. Notably, the \ac{AIMD} had the smallest standard deviation over the year, showing that it adapts very well to different light conditions.

Furthermore, as the number of tags \textit{actively} locating increases, the energy consumption of all tags rises because more \textit{passive} localizations take place. However, this effect remains small for two reasons:
\begin{enumerate*}[label=\textit{\roman*)}]
  \item \textit{active} localization consumes about three times as much energy as \textit{passive} localization, and
  \item most of the additional \textit{active} localizations - especially under the \ac{AIMD} variants, happen during conditions with a higher energy supply.
\end{enumerate*}

Anchors maintain a higher \ac{SoC} than tags, reflecting their lower activity and energy consumption. 
For tags and anchors operating under the low light harvesting profile, the \ac{AIMD}-based algorithm detects the declining battery state according to the metric in~\cref{eq:energyaware-algorithm-metric} and reduces the \textit{active} localization rate to zero. At this point, the devices remain in their lowest sleep state, listening only on the \ac{WuR}. However, even this minimal power consumption cannot be sustained, as the average harvested power of \qty{3.59}{\micro\watt} is not sufficient to maintain even the sleep consumption and self-discharge of the battery. Consequently, all tags and anchors that operated under this lighting profile had completely discharged their batteries already very early in the year.

For anchors, the median \ac{SoC} at the end of the year is between \qtyrange{30}{40}{\percent} across all tested algorithms and network sizes. However, the choice of rate-control algorithm affects how often anchors end up with a low battery (\qtyrange{0}{25}{\percent} \ac{SoC}): with \ac{AIMD}, all anchors fall in this category, while bounded-\ac{AIMD} and constant rate reduce it to about \qty{60}{\percent} and \qty{50}{\percent}, respectively. Increasing the number of tags has only a moderate effect on anchors. Their average \ac{SoC} ranges from \qty{29}{\percent} (with \ac{AIMD}) and \qty{39}{\percent} (constant-rate), showing how \textsc{WakeLoc} effectively stabilizes anchor consumption even as the number of tags scales up.

\cref{fig:energy-aware-localization-results} shows the average number of localizations per day and per tag for the three simulated network sizes (\num{5}, \num{20}, and \num{100} tags). The \ac{AIMD} and bounded-\ac{AIMD} algorithms perform significantly more localizations than the constant-rate algorithm. Especially, this increased activity occurs mainly during the summer months, when above-average amounts of light and thus energy are available. On the strongest days in July, the \ac{AIMD} algorithm achieves up to an average of \num{2400} localizations per day and tag in the \num{100} tags deployment, while the constant-rate algorithm remains constant at around \num{180} localizations (\textit{active} and \textit{passive} combined).

\cref{tab:energy-aware-localization-results} summarizes the statistical parameters of the localizations. For all network sizes, \ac{AIMD} enables the most localizations on average (up to \num{2031}/\num{2554} for tags/anchors), followed by bounded-\ac{AIMD} (\num{761}/\num{811}) and constant-rate (\num{180}/\num{182}). The standard deviations also indicate that the number of localizations depends heavily on daylight availability, which is particularly visible in large networks. Minima of \num{0} indicate that the tags and anchors without energy were unable to perform localization. Maxima of over \num{13000} localizations with \num{100} tags demonstrate the scalability of the \ac{AIMD} approach in times of high energy availability -- always trying to make best use of the available energy. In addition, the distribution of localizations shows that \ac{AIMD} and bounded-\ac{AIMD} respond particularly well to seasonal variation: they make efficient use of phases with plenty of light, while constant-rate remains constant and does not adjust resources. This confirms the advantages of energy-adaptive algorithms, especially in heterogeneous lighting environments where some tags have significantly more energy available than others.

The bounded-\ac{AIMD} algorithm, while achieving a lower average localization rate compared to \ac{AIMD}, shows an advantage during periods of very low light intensity. By conserving energy more aggressively, bounded-\ac{AIMD} maintains higher battery levels, enabling nodes to sustain higher localization rates under scarce energy availability. In contrast, the standard \ac{AIMD} drastically reduces its localization rate in such scenarios, resulting in fewer successful localizations when energy input is minimal.

Summarizing, the \ac{AIMD} achieved in average a \qty{2.7}{\times} to \qty{6.1}{\times} higher localization rate per day and tag then bounded-\ac{AIMD}, and a \qty{11.3}{\times} to \qty{33.7}{\times} higher rate than the constant rate algorithm. At the same time, the battery \ac{SoC} at the end of the year was only \qty{1.6}{\times} to \qty{2.2}{\times}, and \qty{2.9}{\times} to \qty{3.9}{\times}, lower for the tags, respectively. Together, these numbers demonstrate that energy-aware algorithms can significantly improve the localization rate of the whole network, without compromising too much on the battery and almost fulfilling the same design constraints of maintaining an average charge of \qty{7}{\percent} at the end of the year.

Even though the algorithm parameters were tuned under average-light conditions, the scheme scaled well to nodes under more high-light illumination as well: none of the tags or anchors remained with a battery state above \qty{75}{\percent} at the end of the year, indicating efficient utilization of available energy across heterogeneous light conditions.
\section{Conclusion}\label{sec:conclusion}

This work presents \textsc{Eco-WakeLoc}, a scalable and energy-neutral \ac{UWB} indoor localization system that addresses the fundamental trade-off between energy efficiency and responsiveness. By combining event-driven operation with cooperative localization, the system activates anchors only when needed and allows \textit{passive} tags to opportunistically utilize \textit{active} tag localizations, reducing coordination overhead while maintaining centimeter-level accuracy and scalability.

The developed sensor nodes integrate \textsc{WakeMod} \acp{WuR}, \textsc{Qorvo} \textsc{DWM3000} \ac{UWB} transceivers, and indoor solar energy harvesting, achieving ultra-low sleep power of \qty{7.84}{\micro\watt} while remaining idle listening. Controlled experiments and real-world deployment on a quadruped robot confirm robust performance under varying conditions and sub-optimal anchor placement, with average localization errors of \qty{21.89}{\centi\meter} for \textit{active} tags, \qty{25.70}{\centi\meter} for \textit{passive} tags, and \qty{43}{\centi\meter} in realistic dynamic scenarios.

Energy characterization and simulation studies demonstrate the system's sustainability and energy neutrality. \textit{Active} tags consume \qty{3.22}{\milli\joule} per localization, \textit{passive} tags \qty{951}{\micro\joule}, and anchors \qty{353}{\micro\joule}. The introduced energy-aware scheduling algorithm, based on \ac{AIMD} principles, adapts localization rates to the availability of harvested solar energy, enabling up to \num{2031} localizations per tag per day under all conditions while maintaining \qty{7}{\percent} battery capacity on average for the nodes by the end of the simulated year, even when starting from an initial battery level of \qty{0}{\percent}.

\textsc{Eco-WakeLoc} enables maintenance-free, high-accuracy deployments in environments where wired infrastructure is impractical, such as already existing buildings, temporary installations, or large-scale \ac{IoT} networks. Its cooperative and energy-adaptive design demonstrates that real-time indoor localization can achieve both low latency and energy-neutrality, paving the way for sustainable, cost-effective, and flexible deployments in a wide range of applications.

\bibliographystyle{IEEEtran}
\bibliography{bib/references}

@inbook{more78_leven_marquar,
  DATE_ADDED =   {Sat Aug 30 18:38:32 2025},
  author =       {Jorge J. Mor{\'e}},
  booktitle =    {Lecture Notes in Mathematics},
  doi =          {10.1007/bfb0067700},
  pages =        {105-116},
  publisher =    {Springer Berlin Heidelberg},
  series =       {Lecture Notes in Mathematics},
  title =        {The Levenberg-Marquardt algorithm: Implementation
                  and theory},
  url =          {http://dx.doi.org/10.1007/BFb0067700},
  year =         {1978},
  chapter =      {10},
}

@inproceedings{larsson19_optim_trilat_is_eigen_probl,
  title          = {Optimal Trilateration Is an Eigenvalue Problem},
  author         = {Martin Larsson and Viktor Larsson and Kalle Astrom and Magnus Oskarsson},
  year           = 2019,
  month          = 5,
  booktitle      = {ICASSP 2019 - 2019 IEEE International Conference on Acoustics, Speech and Signal Processing (ICASSP)},
  pages          = {5586-5590},
  doi            = {10.1109/icassp.2019.8683355},
  url            = {http://dx.doi.org/10.1109/ICASSP.2019.8683355},
  date_added     = {Mon Jul 15 22:18:10 2024}
}

@article{larsson25_singl_sourc_local_as_eigen_probl,
  author =       {Martin Larsson and Viktor Larsson and Kalle
                  {\AA}str{\"o}m and Magnus Oskarsson},
  title =        {Single-Source Localization As an Eigenvalue Problem},
  journal =      {IEEE Transactions on Signal Processing},
  volume =       73,
  pages =        {574-583},
  year =         2025,
  doi =          {10.1109/tsp.2025.3532102},
  url =          {http://dx.doi.org/10.1109/TSP.2025.3532102},
  DATE_ADDED =   {Sat Aug 30 18:53:32 2025},
}

@article{marquardt63_algor_least_squar_estim_nonlin_param,
  author =       {Donald W. Marquardt},
  title =        {An Algorithm for Least-Squares Estimation of
                  Nonlinear Parameters},
  journal =      {Journal of the Society for Industrial and Applied
                  Mathematics},
  volume =       11,
  number =       2,
  pages =        {431-441},
  year =         1963,
  doi =          {10.1137/0111030},
  url =          {http://dx.doi.org/10.1137/0111030},
  DATE_ADDED =   {Sat Aug 30 18:59:03 2025},
}

@article{ding25_optim_placem_minim_gdop_with,
  author =       {Yanwu Ding and Dan Shen and Khanh Pham and Genshe
                  Chen},
  title =        {Optimal Placements for Minimum Gdop With
                  Consideration on the Elevations of Access Nodes},
  journal =      {IEEE Transactions on Instrumentation and
                  Measurement},
  volume =       74,
  pages =        {1-10},
  year =         2025,
  doi =          {10.1109/tim.2024.3497055},
  url =          {http://dx.doi.org/10.1109/TIM.2024.3497055},
  DATE_ADDED =   {Sat Aug 30 19:03:35 2025},
}

@inproceedings{schulthess25_wakem,
  author =       {Lukas Schulthess and Silvano Cortesi and Michele
                  Magno},
  title =        {WakeMod: A 6.9 $\mu$W Wake-Up Radio Module with
                  -72.6 dBm Sensitivity for On-Demand IoT},
  booktitle =    {2025 10th International Workshop on Advances in
                  Sensors and Interfaces (IWASI)},
  year =         2025,
  pages =        {1-6},
  doi =          {10.1109/iwasi66786.2025.11122000},
  url =          {http://dx.doi.org/10.1109/IWASI66786.2025.11122000},
  DATE_ADDED =   {Mon Aug 25 08:15:32 2025},
  month =        7,
}

@article{bouazzaoui22_enhan_rgb_d_slam_perfor,
  author =       {Imad El Bouazzaoui and Sergio A. Rodriguez Florez
                  and Abdelhafid El Ouardi},
  title =        {Enhancing Rgb-D Slam Performances Considering Sensor
                  Specifications for Indoor Localization},
  journal =      {IEEE Sensors Journal},
  volume =       22,
  number =       6,
  pages =        {4970-4977},
  year =         2022,
  doi =          {10.1109/jsen.2021.3073676},
  url =          {http://dx.doi.org/10.1109/JSEN.2021.3073676},
  DATE_ADDED =   {Thu Sep 4 18:00:24 2025},
}

@article{hayward22_survey_indoor_locat_techn_techn_applic_indus,
  author =       {S.J. Hayward and K. van Lopik and C. Hinde and
                  A.A. West},
  title =        {A Survey of Indoor Location Technologies, Techniques
                  and Applications in Industry},
  journal =      {Internet of Things},
  volume =       20,
  pages =        100608,
  year =         2022,
  doi =          {10.1016/j.iot.2022.100608},
  url =          {http://dx.doi.org/10.1016/j.iot.2022.100608},
  DATE_ADDED =   {Thu Sep 4 18:02:45 2025},
}

@article{mayer24_self_sustain_ultraw_posit_system,
  title          = {Self-Sustaining Ultrawideband Positioning System for Event-Driven Indoor Localization},
  author         = {Philipp Mayer and Michele Magno and Luca Benini},
  year           = 2024,
  journal        = {IEEE Internet of Things Journal},
  volume         = 11,
  number         = 1,
  pages          = {1272-1284},
  doi            = {10.1109/jiot.2023.3289568},
  url            = {http://dx.doi.org/10.1109/JIOT.2023.3289568},
  date_added     = {Wed May 15 11:43:28 2024}
}

@article{patru23_flext,
  title          = {Flextdoa: Robust and Scalable Time-Difference of Arrival Localization Using Ultra-Wideband Devices},
  author         = {George-Cristian Pătru and Laura Flueratoru and Iuliu Vasilescu and Drago{\c{s}} Niculescu and Daniel Rosner},
  year           = 2023,
  journal        = {IEEE Access},
  volume         = 11,
  pages          = {28610-28627},
  doi            = {10.1109/access.2023.3259320},
  url            = {http://dx.doi.org/10.1109/ACCESS.2023.3259320},
  date_added     = {Wed May 15 10:32:24 2024}
}

@article{li25_indoor_uwb_local_method_based,
  author =       {Yaning Li and Baoguo Yu and Lu Huang},
  title =        {An Indoor Uwb Localization Method Based on Adaptive
                  Channel Bias Estimation},
  journal =      {IEEE Sensors Journal},
  volume =       25,
  number =       1,
  pages =        {1339-1349},
  year =         2025,
  doi =          {10.1109/jsen.2024.3493069},
  url =          {http://dx.doi.org/10.1109/JSEN.2024.3493069},
  DATE_ADDED =   {Thu Sep 4 19:23:49 2025},
}

@article{istomin21_janus,
  author =       {Timofei Istomin and Elia Leoni and Davide Molteni
                  and Amy L. Murphy and Gian Pietro Picco and Maurizio
                  Griva},
  title =        {Janus},
  journal =      {Proceedings of the ACM on Interactive, Mobile,
                  Wearable and Ubiquitous Technologies},
  volume =       5,
  number =       4,
  pages =        {1-33},
  year =         2021,
  doi =          {10.1145/3494978},
  url =          {http://dx.doi.org/10.1145/3494978},
  DATE_ADDED =   {Thu Sep 4 19:29:45 2025},
}

@inproceedings{polonelli22_perfor_compar_decaw_dw100_dw300,
  title          = {Performance Comparison between Decawave DW1000 and DW3000 in low-power double side ranging applications},
  author         = {Tommaso Polonelli and Simon Schlapfer and Michele Magno},
  year           = 2022,
  month          = 8,
  booktitle      = {2022 IEEE Sensors Applications Symposium (SAS)},
  doi            = {10.1109/sas54819.2022.9881375},
  url            = {http://dx.doi.org/10.1109/sas54819.2022.9881375},
  date_added     = {Wed Mar 22 09:27:50 2023}
}

@inproceedings{luder25_anitr,
  author =       {Victor Luder and Lukas Schulthess and Silvano
                  Cortesi and Leyla Rivero Davis and Michele Magno},
  title =        {AniTrack: A Power-Efficient, Time-Slotted and Robust
                  UWB Localization System for Animal Tracking in a
                  Controlled Setting},
  booktitle =    {2025 10th International Workshop on Advances in
                  Sensors and Interfaces (IWASI)},
  year =         2025,
  pages =        {1-6},
  doi =          {10.1109/iwasi66786.2025.11121986},
  url =          {http://dx.doi.org/10.1109/IWASI66786.2025.11121986},
  DATE_ADDED =   {Mon Aug 25 08:15:57 2025},
  month =        7,
}

@article{laadung22_novel_activ_passiv_two_way,
  title          = {Novel Active-Passive Two-Way Ranging Protocols for Uwb Positioning Systems},
  author         = {Taavi Laadung and Sander Ulp and Muhammad Mahtab Alam and Yannick Le Moullec},
  year           = 2022,
  journal        = {IEEE Sensors Journal},
  volume         = 22,
  number         = 6,
  pages          = {5223--5237},
  doi            = {10.1109/jsen.2021.3125570},
  url            = {http://dx.doi.org/10.1109/JSEN.2021.3125570},
  date_added     = {Thu Jul 18 22:59:14 2024}
}

@article{zhao21_uloc,
  author =       {Minghui Zhao and Tyler Chang and Aditya Arun and
                  Roshan Ayyalasomayajula and Chi Zhang and Dinesh
                  Bharadia},
  title =        {Uloc},
  journal =      {Proceedings of the ACM on Interactive, Mobile,
                  Wearable and Ubiquitous Technologies},
  volume =       5,
  number =       3,
  pages =        {1-31},
  year =         2021,
  doi =          {10.1145/3478124},
  url =          {http://dx.doi.org/10.1145/3478124},
  DATE_ADDED =   {Thu Apr 10 15:11:10 2025},
}

@article{ahmed19_optim_power_manag_with_guaran,
  author =       {Rehan Ahmed and Bernhard Buchli and Stefan Draskovic
                  and Lukas Sigrist and Pratyush Kumar and Lothar
                  Thiele},
  title =        {Optimal Power Management With Guaranteed Minimum
                  Energy Utilization for Solar Energy Harvesting
                  Systems},
  journal =      {ACM Transactions on Embedded Computing Systems},
  volume =       18,
  number =       4,
  pages =        {1-26},
  year =         2019,
  doi =          {10.1145/3317679},
  url =          {http://dx.doi.org/10.1145/3317679},
  DATE_ADDED =   {Fri Sep 5 09:08:17 2025},
}

@inproceedings{cortesi25_wakel,
  author =       {Silvano Cortesi and Christian Vogt and Michele
                  Magno},
  title =        {WakeLoc: An Ultra-Low Power, Accurate and Scalable
                  On-Demand RTLS using Wake-Up Radios},
  booktitle =    {IEEE INFOCOM 2025 - IEEE Conference on Computer
                  Communications Workshops (INFOCOM WKSHPS)},
  year =         2025,
  pages =        {1-6},
  doi =          {10.1109/infocomwkshps65812.2025.11152965},
  url =
                  {http://dx.doi.org/10.1109/INFOCOMWKSHPS65812.2025.11152965},
  DATE_ADDED =   {Wed Sep 17 18:36:07 2025},
  month =        5,
}

@article{verma22_novel_rf_energ_harves_event,
  author =       {Gourav Verma and Vidushi Sharma},
  title =        {A Novel Rf Energy Harvester for Event-Based
                  Environmental Monitoring in Wireless Sensor
                  Networks},
  journal =      {IEEE Internet of Things Journal},
  volume =       9,
  number =       5,
  pages =        {3189-3203},
  year =         2022,
  doi =          {10.1109/jiot.2021.3097629},
  url =          {http://dx.doi.org/10.1109/JIOT.2021.3097629},
  DATE_ADDED =   {Fri Sep 5 09:13:29 2025},
}

@inproceedings{villani24_ultra_wideb_wake_up_receiv,
  title          = {A 36nW Ultra-Wideband Wake-Up Receiver with -86dBm Sensitivity and Addressing Capabilities},
  author         = {Federico Villani and Enea Masina and Thomas Burger and Michele Magno},
  year           = 2024,
  month          = 5,
  booktitle      = {2024 IEEE International Symposium on Circuits and Systems (ISCAS)},
  pages          = 5,
  doi            = {10.1109/iscas58744.2024.10558556},
  url            = {http://dx.doi.org/10.1109/ISCAS58744.2024.10558556},
  date_added     = {Wed Jul 24 14:40:43 2024}
}

@inproceedings{kazdaridis21_a_novel_archit,
  author =       {Giannis Kazdaridis and Nikos Sidiropoulos and
                  Ioannis Zografopoulos and Thanasis Korakis},
  title =        {A Novel Architecture for Semi-Active Wake-Up
                  Radios Attaining Sensitivity Beyond -70dBm: Demo Abstrace},
  booktitle =    {2021 20th International Conference on Information Processing in Sensor Networks},
  year =         2021,
  pages =        {398-399},
  doi =          {10.1145/3412382.3458782},
  url =          {https://dx.doi.org/10.1145/3412382.3458782},
  DATE_ADDED =   {Fri Mar 21 13:11:20 2025},
  month =        5,
}

@article{balsamo16_hiber,
  author =       {Domenico Balsamo and Alex S. Weddell and Anup Das
                  and Alberto Rodriguez Arreola and Davide Brunelli
                  and Bashir M. Al-Hashimi and Geoff V. Merrett and
                  Luca Benini},
  title =        {Hibernus++: a Self-Calibrating and Adaptive System
                  for Transiently-Powered Embedded Devices},
  journal =      {IEEE Transactions on Computer-Aided Design of
                  Integrated Circuits and Systems},
  volume =       35,
  number =       12,
  pages =        {1968-1980},
  year =         2016,
  doi =          {10.1109/tcad.2016.2547919},
  url =          {http://dx.doi.org/10.1109/TCAD.2016.2547919},
  DATE_ADDED =   {Fri Sep 5 09:18:26 2025},
}

@inproceedings{giouroukis20,
  author =       {Dimitrios Giouroukis and Alexander Dadiani and Jonas
                  Traub and Steffen Zeuch and Volker Markl},
  title =        {A survey of adaptive sampling and filtering
                  algorithms for the internet of things},
  booktitle =    {Proceedings of the 14th ACM International Conference
                  on Distributed and Event-based Systems},
  year =         2020,
  pages =        {27-38},
  doi =          {10.1145/3401025.3403777},
  url =          {http://dx.doi.org/10.1145/3401025.3403777},
  DATE_ADDED =   {Fri Sep 5 09:20:22 2025},
  month =        7,
}

@inproceedings{giordano23_energ_aware_adapt_sampl_self,
  title          = {Energy-Aware Adaptive Sampling for Self-Sustainability in Resource-Constrained IoT Devices},
  author         = {Marco Giordano and Silvano Cortesi and Prodromos-Vasileios Mekikis and Michele Crabolu and Giovanni Bellusci and Michele Magno},
  year           = 2023,
  month          = 11,
  booktitle      = {Proceedings of the 11th International Workshop on Energy Harvesting \& Energy-Neutral Sensing Systems},
  doi            = {10.1145/3628353.3628545},
  url            = {http://dx.doi.org/10.1145/3628353.3628545},
  date_added     = {Thu Jan 25 15:08:57 2024}
}

@inproceedings{dotlic18_rangin_method_utiliz_carrier_frequen_offset_estim,
  title          = {Ranging Methods Utilizing Carrier Frequency Offset Estimation},
  author         = {Igor Dotlic and Andrew Connell and Michael McLaughlin},
  year           = 2018,
  month          = 10,
  booktitle      = {2018 15th Workshop on Positioning, Navigation and Communications (WPNC)},
  pages          = 6,
  doi            = {10.1109/wpnc.2018.8555809},
  url            = {http://dx.doi.org/10.1109/WPNC.2018.8555809},
  date_added     = {Mon Jul 15 16:51:52 2024}
}

@article{coppens22_overv_uwb_stand_organ_ieee,
  title          = {An Overview of Uwb Standards and Organizations (IEEE 802.15.4, Fira, Apple): Interoperability Aspects and Future Research Directions},
  author         = {Dieter Coppens and Adnan Shahid and Sam Lemey and Ben Van Herbruggen and Chris Marshall and Eli De Poorter},
  year           = 2022,
  journal        = {IEEE Access},
  volume         = 10,
  pages          = {70219--70241},
  doi            = {10.1109/access.2022.3187410},
  url            = {http://dx.doi.org/10.1109/ACCESS.2022.3187410},
  date_added     = {Tue Mar 21 15:38:52 2023}
}

@article{yang22_high_precis_uwb_based_local,
  author =       {Beiya Yang and Erfu Yang and Leijian Yu and Andrew
                  Loeliger},
  title =        {High-Precision Uwb-Based Localisation for Uav in
                  Extremely Confined Environments},
  journal =      {IEEE Sensors Journal},
  volume =       22,
  number =       1,
  pages =        {1020-1029},
  year =         2022,
  doi =          {10.1109/jsen.2021.3130724},
  url =          {http://dx.doi.org/10.1109/JSEN.2021.3130724},
  DATE_ADDED =   {Fri Sep 5 12:41:47 2025},
}

@misc{bilge22_evaluat_ultra_wide_band_techn,
  title          = {Evaluation of Ultra Wide Band Technology As an Enhancement for Ble Based Location Estimation},
  author         = {Miro Bilge},
  year           = 2022,
  publisher =    {arXiv},
  doi            = {10.48550/ARXIV.2202.00558},
  url            = {https://arxiv.org/abs/2202.00558},
  date_added     = {Wed Mar 22 09:28:08 2023}
}

@misc{ramesh20_robus_scalab_techn_twr_tdoa,
  title          = {Robust and Scalable Techniques for Twr and Tdoa Based Localization Using Ultra Wide Band Radios},
  author         = {Rakshit Ramesh and Aaron John-Sabu and Harshitha S and Siddarth Ramesh and Vishwas Navada B and Mukunth Arunachalam and Bharadwaj Amrutur},
  year           = 2020,
  publisher =    {arXiv},
  pages          = 6,
  doi            = {10.48550/ARXIV.2008.04248},
  url            = {https://arxiv.org/abs/2008.04248},
  date_added     = {Tue Jul 30 17:18:49 2024}
}

@article{ridolfi18_analy_scalab_uwb_indoor_local,
  title          = {Analysis of the Scalability of Uwb Indoor Localization Solutions for High User Densities},
  author         = {Matteo Ridolfi and Samuel Van de Velde and Heidi Steendam and Eli De Poorter},
  year           = 2018,
  journal        = {Sensors},
  volume         = 18,
  number         = 6,
  pages          = 1875,
  doi            = {10.3390/s18061875},
  url            = {http://dx.doi.org/10.3390/s18061875},
  date_added     = {Tue Jul 30 17:22:28 2024}
}

@article{yang22_vuloc,
  title          = {Vuloc},
  author         = {Jing Yang and BaiShun Dong and Jiliang Wang},
  year           = 2022,
  journal        = {Proceedings of the ACM on Interactive, Mobile, Wearable and Ubiquitous Technologies},
  volume         = 6,
  number         = 3,
  pages          = {1--25},
  doi            = {10.1145/3550286},
  url            = {http://dx.doi.org/10.1145/3550286},
  date_added     = {Tue Jul 30 17:32:49 2024}
}

@article{corbalan20_ultra_wideb_concur_rangin,
  title          = {Ultra-Wideband Concurrent Ranging},
  author         = {Pablo Corbal{\'a}n and Gian Pietro Picco},
  year           = 2020,
  journal        = {ACM Transactions on Sensor Networks},
  volume         = 16,
  number         = 4,
  pages          = {1--41},
  doi            = {10.1145/3409477},
  url            = {http://dx.doi.org/10.1145/3409477},
  date_added     = {Thu Jun 13 07:04:58 2024}
}

@inproceedings{santoro21_scale,
  title          = {Scale up to infinity: the UWB Indoor Global Positioning System},
  author         = {Luca Santoro and Matteo Nardello and Davide Brunelli and Daniele Fontanelli},
  year           = 2021,
  month          = 10,
  booktitle      = {2021 IEEE International Symposium on Robotic and Sensors Environments (ROSE)},
  pages          = 8,
  doi            = {10.1109/rose52750.2021.9611770},
  url            = {http://dx.doi.org/10.1109/ROSE52750.2021.9611770},
  date_added     = {Tue Jul 30 18:03:44 2024}
}

@inproceedings{spirito98_hyperbolic_posi,
  title          = {On the hyperbolic positioning of GSM mobile stations},
  author         = {M.A. Spirito and A.G. Mattioli},
  year           = 1998,
  month          = 10,
  booktitle      = {1998 URSI International Symposium on Signals, Systems, and Electronics. Conference Proceedings (Cat. No.98EX167)},
  pages          = 5,
  doi            = {10.1109/issse.1998.738060},
  url            = {http://dx.doi.org/10.1109/ISSSE.1998.738060},
  date_added     = {Tue Jul 30 18:21:37 2024}
}

@inproceedings{gust03positioning_tdoa,
  title          = {Positioning using time-difference of arrival measurements},
  author         = {F. Gustafsson and F. Gunnarsson},
  year           = 2003,
  month          = {-},
  booktitle      = {2003 IEEE International Conference on Acoustics, Speech, and Signal Processing, 2003. Proceedings. (ICASSP '03).},
  doi            = {10.1109/icassp.2003.1201741},
  url            = {http://dx.doi.org/10.1109/ICASSP.2003.1201741},
  date_added     = {Tue Nov 7 15:26:10 2023}
}

@misc{renata_userguide,
  title =        {Rechargeable lithium-ion polymer batteries guidelines},
  url =          {https://www.renata.com/en/downloadfile/renata-rechargeable-lipo-guidelines/?fileid=f871fff174d88fe997723e20ed},
  author =       {renata},
  note =         {Accessed: 05.09.2025}
}

@inproceedings{grosswindhager19_snapl,
  author =       {Bernhard Gro{\ss}windhager and Michael Stocker and
                  Michael Rath and Carlo Alberto Boano and Kay
                  R{\"o}mer},
  title =        {SnapLoc},
  booktitle =    {Proceedings of the 18th International Conference on
                  Information Processing in Sensor Networks},
  year =         2019,
  doi =          {10.1145/3302506.3310389},
  url =          {http://dx.doi.org/10.1145/3302506.3310389},
  DATE_ADDED =   {Tue Jul 30 17:25:38 2024},
  month =        4,
}

@article{herbruggen24_real_time_anchor_node_selec,
  author =       {Ben Van Herbruggen and Dries Van Leemput and Jaro
                  Van Landschoot and Eli De Poorter},
  title =        {Real-Time Anchor Node Selection for Two-Way-Ranging
                  (TWR) Ultra-Wideband (UWB) Indoor Positioning
                  Systems},
  journal =      {IEEE Sensors Letters},
  volume =       8,
  number =       3,
  pages =        {1-4},
  year =         2024,
  doi =          {10.1109/lsens.2024.3363231},
  url =          {http://dx.doi.org/10.1109/LSENS.2024.3363231},
  DATE_ADDED =   {Tue Jun 11 19:34:39 2024},
}

@inproceedings{sigrist19_datas,
  author =       {Lukas Sigrist and Andres Gomez and Lothar Thiele},
  title =        {Dataset: Tracing Indoor Solar Harvesting},
  booktitle =    {Proceedings of the 2nd Workshop on Data Acquisition
                  To Analysis},
  year =         2019,
  pages =        {47-50},
  doi =          {10.1145/3359427.3361910},
  url =          {http://dx.doi.org/10.1145/3359427.3361910},
  DATE_ADDED =   {Sat Sep 6 22:19:05 2025},
  month =        11,
}

@inproceedings{molteni22_cloves,
  title          = {Cloves},
  author         = {Davide Molteni and Gian Pietro Picco and Matteo Trobinger and Davide Vecchia},
  year           = 2022,
  month          = 11,
  booktitle      = {Proceedings of the 20th ACM Conference on Embedded Networked Sensor Systems},
  pages          = {808-809},
  doi            = {10.1145/3560905.3568072},
  url            = {http://dx.doi.org/10.1145/3560905.3568072},
  date_added     = {Tue Jul 23 00:02:07 2024}
}

@inproceedings{kolakowski16_tdoa_twr_uwb,
  author =       {Marcin Kolakowski and Vitomir Djaja-Josko},
  title =        {TDOA-TWR based positioning algorithm for UWB
                  localization system},
  booktitle =    {2016 21st International Conference on Microwave,
                  Radar and Wireless Communications (MIKON)},
  year =         2016,
  pages =        {1-4},
  doi =          {10.1109/mikon.2016.7491981},
  url =          {http://dx.doi.org/10.1109/MIKON.2016.7491981},
  DATE_ADDED =   {Thu Sep 11 10:09:53 2025},
  month =        5,
}

@article{piyare17_ultra_low_power_wake_up_radios,
  title          = {Ultra Low Power Wake-Up Radios: a Hardware and Networking Survey},
  author         = {Rajeev Piyare and Amy L. Murphy and Csaba Kiraly and Pietro Tosato and Davide Brunelli},
  year           = 2017,
  journal        = {IEEE Communications Surveys \& Tutorials},
  volume         = 19,
  number         = 4,
  pages          = {2117--2157},
  doi            = {10.1109/comst.2017.2728092},
  url            = {http://dx.doi.org/10.1109/COMST.2017.2728092},
  date_added     = {Fri Apr 26 14:42:20 2024}
}

@inproceedings{polonelli21_ultra_low_power_wake_up,
  title          = {Ultra-Low Power Wake-Up Receiver for Location Aware Objects Operating with UWB},
  author         = {Tommaso Polonelli and Federico Villani and Michele Magno},
  year           = 2021,
  month          = 10,
  booktitle      = {2021 17th International Conference on Wireless and Mobile Computing, Networking and Communications (WiMob)},
  pages          = 8,
  doi            = {10.1109/wimob52687.2021.9606248},
  url            = {http://dx.doi.org/10.1109/WiMob52687.2021.9606248},
  date_added     = {Fri Feb 16 15:45:25 2024}
}

@inproceedings{sutton15_zippy,
  title          = {Zippy},
  author         = {Felix Sutton and Bernhard Buchli and Jan Beutel and Lothar Thiele},
  year           = 2015,
  month          = 11,
  booktitle      = {Proceedings of the 13th ACM Conference on Embedded Networked Sensor Systems},
  doi            = {10.1145/2809695.2809705},
  url            = {http://dx.doi.org/10.1145/2809695.2809705},
  date_added     = {Wed Mar 22 09:22:39 2023}
}

@article{beck08_exact_approx_solut_sourc_local_probl,
  author =       {A. Beck and P. Stoica and Jian Li},
  title =        {Exact and Approximate Solutions of Source
                  Localization Problems},
  journal =      {IEEE Transactions on Signal Processing},
  volume =       56,
  number =       5,
  pages =        {1770-1778},
  year =         2008,
  doi =          {10.1109/tsp.2007.909342},
  url =          {http://dx.doi.org/10.1109/TSP.2007.909342},
  DATE_ADDED =   {Thu Sep 11 16:31:17 2025},
}

@article{luke17_simpl_global_conver_algor_nonsm,
  author =       {D. Russell Luke and Shoham Sabach and Marc Teboulle
                  and Kobi Zatlawey},
  title =        {A Simple Globally Convergent Algorithm for the
                  Nonsmooth Nonconvex Single Source Localization
                  Problem},
  journal =      {Journal of Global Optimization},
  volume =       69,
  number =       4,
  pages =        {889-909},
  year =         2017,
  doi =          {10.1007/s10898-017-0545-6},
  url =          {http://dx.doi.org/10.1007/s10898-017-0545-6},
  DATE_ADDED =   {Thu Sep 11 16:31:25 2025},
}

@article{zhou10_closed_form_algor_least_squar_trilat_probl,
  author =       {Yu Zhou},
  title =        {A Closed-Form Algorithm for the Least-Squares
                  Trilateration Problem},
  journal =      {Robotica},
  volume =       29,
  number =       3,
  pages =        {375-389},
  year =         2010,
  doi =          {10.1017/s0263574710000196},
  url =          {http://dx.doi.org/10.1017/S0263574710000196},
  DATE_ADDED =   {Thu Sep 11 16:31:34 2025},
}

@inproceedings{zandian18_nlos_detec_mitig_differ_local,
  author =       {Reza Zandian and Ulf Witkowski},
  title =        {NLOS Detection and Mitigation in Differential
                  Localization Topologies Based on UWB Devices},
  booktitle =    {2018 International Conference on Indoor Positioning
                  and Indoor Navigation (IPIN)},
  year =         2018,
  pages =        {1-8},
  doi =          {10.1109/ipin.2018.8533781},
  url =          {http://dx.doi.org/10.1109/IPIN.2018.8533781},
  DATE_ADDED =   {Mon Sep 15 10:45:25 2025},
  month =        9,
}

@inproceedings{chen25_low_area_wake_up_receiv,
  author =       {Binbin Chen and Heyu Ren and Wenjun Gong and
                  Liangjian Lyu and C.-J.Richard Shi},
  title =        {A 915MHz 97nW Low-Area Wake-Up Receiver with an
                  Envelope-Tracking Mixer Achieving -73.2dBm
                  Sensitivity},
  booktitle =    {2025 IEEE International Symposium on Circuits and
                  Systems (ISCAS)},
  year =         2025,
  pages =        {1-5},
  doi =          {10.1109/iscas56072.2025.11043883},
  url =          {http://dx.doi.org/10.1109/ISCAS56072.2025.11043883},
  DATE_ADDED =   {Fri Sep 19 20:38:24 2025},
  month =        5,
}

@misc{polymaps, 
  title =        {PolyMaps indoor map application},
  url =          {https://ethz.ch/staffnet/en/organisation/departments/engineering-and-systems/polymaps.html},
  author =       {ETH Zürich},
  note =         {Accessed: 20.09.2025}
}

@inproceedings{plozza24_auton_navig_dynam_human_envir,
  author =       {Davide Plozza and Steven Marty and Cyril Scherrer
                  and Simon Schwartz and Stefan Zihlmann and Michele
                  Magno},
  title =        {Autonomous Navigation in Dynamic Human Environments
                  with an Embedded 2D LiDAR-based Person Tracker},
  booktitle =    {2024 IEEE Sensors Applications Symposium (SAS)},
  year =         2024,
  pages =        {1-6},
  doi =          {10.1109/sas60918.2024.10636369},
  url =          {http://dx.doi.org/10.1109/SAS60918.2024.10636369},
  DATE_ADDED =   {Wed Sep 24 19:11:49 2025},
  month =        7,
}

@article{chen24_centim_level_indoor_posit_with,
  author =       {Lien-Wu Chen and Chi-Ren Chen},
  title =        {Centimeter-Level Indoor Positioning With Facing
                  Direction Detection for Microlocation-Aware
                  Services},
  journal =      {IEEE Transactions on Systems, Man, and Cybernetics:
                  Systems},
  volume =       54,
  number =       10,
  pages =        {6458-6468},
  year =         2024,
  doi =          {10.1109/tsmc.2024.3432615},
  url =          {http://dx.doi.org/10.1109/TSMC.2024.3432615},
  DATE_ADDED =   {Tue Sep 30 14:32:54 2025},
}

@article{silva22_real_world_deploy_low_cost,
  author =       {Ivo Silva and Cristiano Pendao and Adriano Moreira},
  title =        {Real-World Deployment of Low-Cost Indoor Positioning
                  Systems for Industrial Applications},
  journal =      {IEEE Sensors Journal},
  volume =       22,
  number =       6,
  pages =        {5386-5397},
  year =         2022,
  doi =          {10.1109/jsen.2021.3103662},
  url =          {http://dx.doi.org/10.1109/JSEN.2021.3103662},
  DATE_ADDED =   {Tue Sep 30 14:34:50 2025},
}

@article{qi24_calib_compen_anchor_posit_uwb_indoor_local,
  author =       {Mingyang Qi and Bing Xue and Wei Wang},
  title =        {Calibration and Compensation of Anchor Positions for
                  Uwb Indoor Localization},
  journal =      {IEEE Sensors Journal},
  volume =       24,
  number =       1,
  pages =        {689-699},
  year =         2024,
  doi =          {10.1109/jsen.2023.3329535},
  url =          {http://dx.doi.org/10.1109/JSEN.2023.3329535},
  DATE_ADDED =   {Tue Sep 30 14:36:56 2025},
}

@article{sayfoori25_high_precis_uwb_sensor_based,
  author =       {Reza Sayfoori and Mao-Hsiang Huang and Amir Naderi
                  and Mehwish Bhatti and Ron D. Frostig and Hung Cao},
  title =        {High-Precision Uwb Sensor-Based Real-Time Locating
                  System for Rodent Behavioral Studies},
  journal =      {IEEE Sensors Journal},
  volume =       25,
  number =       13,
  pages =        {25551-25559},
  year =         2025,
  doi =          {10.1109/jsen.2025.3572031},
  url =          {http://dx.doi.org/10.1109/JSEN.2025.3572031},
  DATE_ADDED =   {Tue Sep 30 14:39:07 2025},
}

@inproceedings{venkatapathy15_phynode,
  author={Ramachandran Venkatapathy, Aswin Karthik and Riesner, Andreas and Roidl, Moritz and Emmerich, Jan and Hompel, Michael ten},
  title={PhyNode: An intelligent, cyber-physical system with energy neutral operation for PhyNetLab}, 
  booktitle={Smart SysTech 2015; European Conference on Smart Objects, Systems and Technologies}, 
  year={2015},
  pages={1-8},
  DATE_ADDED = {Tue Sep 30 19:06:00 2025},
}

@article{chen24_room_level_indoor_local_using,
  author =       {Yutao Chen and Yun Wang and Yubin Zhao},
  title =        {A Room-Level Indoor Localization Using an
                  Energy-Harvesting Ble Tag},
  journal =      {Electronics},
  volume =       13,
  number =       22,
  pages =        4493,
  year =         2024,
  doi =          {10.3390/electronics13224493},
  url =          {http://dx.doi.org/10.3390/electronics13224493},
  DATE_ADDED =   {Tue Sep 30 19:12:30 2025},
}

@article{sullivan23_total_cost_owner_real_time,
  title          = {Total Cost of Ownership of Real-Time Locating System (RTLS) Technologies in Factories},
  author         = {B. Patrick Sullivan and Poorya Ghafoorpoor Yazdi and Sebastian Thiede},
  year           = 2023,
  journal        = {Procedia CIRP},
  volume         = 120,
  pages          = {822--827},
  doi            = {10.1016/j.procir.2023.09.082},
  url            = {http://dx.doi.org/10.1016/j.procir.2023.09.082},
  date_added     = {Thu Aug 1 01:47:47 2024}
}

@article{boulos12_real_time_locat_system_rtls_healt,
  title          = {Real-Time Locating Systems (RTLS) in Healthcare: a Condensed Primer},
  author         = {Maged N Kamel Boulos and Geoff Berry},
  year           = 2012,
  journal        = {International Journal of Health Geographics},
  volume         = 11,
  number         = 1,
  pages          = 25,
  doi            = {10.1186/1476-072x-11-25},
  url            = {https://doi.org/10.1186/1476-072x-11-25},
  date_added     = {Fri Sep 10 17:16:18 2021}
}

@article{zhang25_data_set_uwb_cooper_navig,
  title          = {Data Set for Uwb Cooperative Navigation and Positioning of Uav Cluster},
  author         = {Cunle Zhang and Chengkai Tang and Haonan Wang and Baowang Lian and Lingling Zhang},
  year           = 2025,
  journal        = {Scientific Data},
  publisher      = {Nature},
  volume         = 12,
  number         = 1,
  pages          = 486,
  doi            = {10.1038/s41597-025-04808-0},
  url            = {http://dx.doi.org/10.1038/s41597-025-04808-0},
  date_added     = {Wed Nov 19 17:33:55 2025}
}

@article{malivert23_compar_improv_multil_solvin_simul,
  author =       {Franck Malivert and Ouiddad Labbani-Igbida and
                  Herv{\'e} Boeglen},
  title =        {Comparison and Improvement of 3d-Multilateration for
                  Solving Simultaneous Localization of Drones and Uwb
                  Anchors},
  journal =      {Applied Sciences},
  volume =       13,
  number =       2,
  pages =        1002,
  year =         2023,
  doi =          {10.3390/app13021002},
  url =          {http://dx.doi.org/10.3390/app13021002},
  DATE_ADDED =   {Fri Dec 5 09:15:46 2025},
}

@inproceedings{frisch25_why_not_use_tdoa,
  address = {College Station, Texas},
  author = {Daniel Frisch and Uwe D. Hanebeck},
  booktitle = {Proceedings of the 2025 IEEE International Conference on Multisensor Fusion and Integration for Intelligent Systems (MFI 2025)},
  month = {September},
  title = {Why You Shouldn't Use TDOA for Multilateration},
  year = {2025},
  url = {https://isas.iar.kit.edu/pdf/MFI25_Frisch.pdf},
  note = {Accessed: 08.12.2025},
  DATE_ADDED =   {Fri Dec 5 09:15:46 2025},
}

@inproceedings{jiang24_demo,
  author =       {Fan Jiang and Ashutosh Dhekne},
  title =        {Demo: uFi$\mu$: An open-source integrated
                  UWB-WiFi-IMU platform for localization research and
                  beyond},
  booktitle =    {Proceedings of the 25th International Workshop on
                  Mobile Computing Systems and Applications},
  year =         2024,
  pages =        {156-156},
  doi =          {10.1145/3638550.3643628},
  url =          {http://dx.doi.org/10.1145/3638550.3643628},
  DATE_ADDED =   {Tue Jan 6 10:14:30 2026},
  month =        2,
}
\vfill
\pagebreak
\begin{IEEEbiography}[{\includegraphics[width=1in,height=1.25in,clip,keepaspectratio]{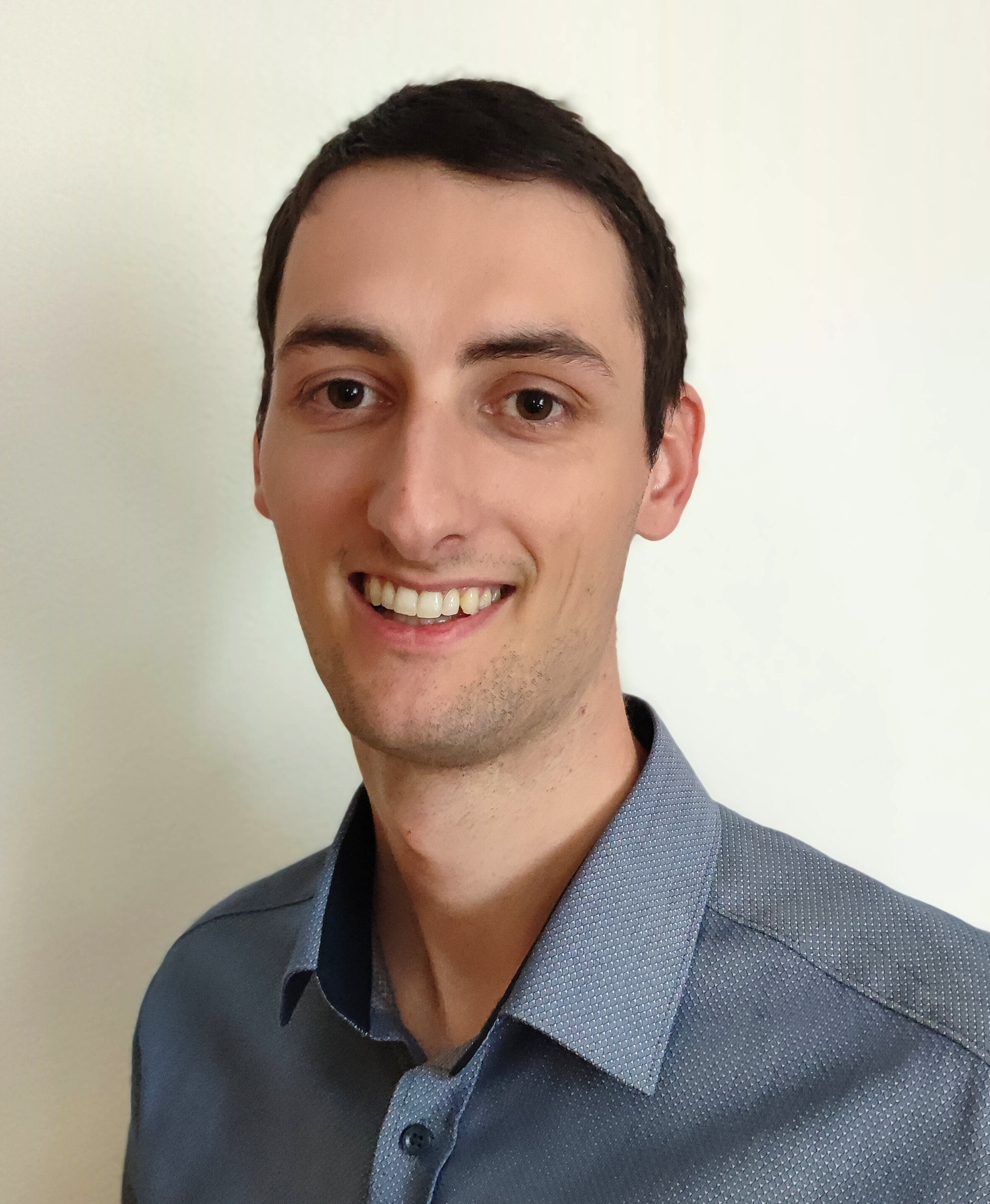}}]{Silvano Cortesi} (GS'22) received the B. Sc. and the M.Sc. degrees in electronics engineering and information technology from ETH Zürich, Zürich, Switzerland in 2020 and 2021, respectively. He is currently pursuing his Ph.D. degree with the Center for Project-Based Learning at ETH Zürich, Zürich, Switzerland. His research work focuses on indoor localization, ultra-low power and self-sustainable IoT, wireless sensor networks and energy harvesting.
\end{IEEEbiography}

\begin{IEEEbiography}[{\includegraphics[width=1in,height=1.25in,clip,keepaspectratio]{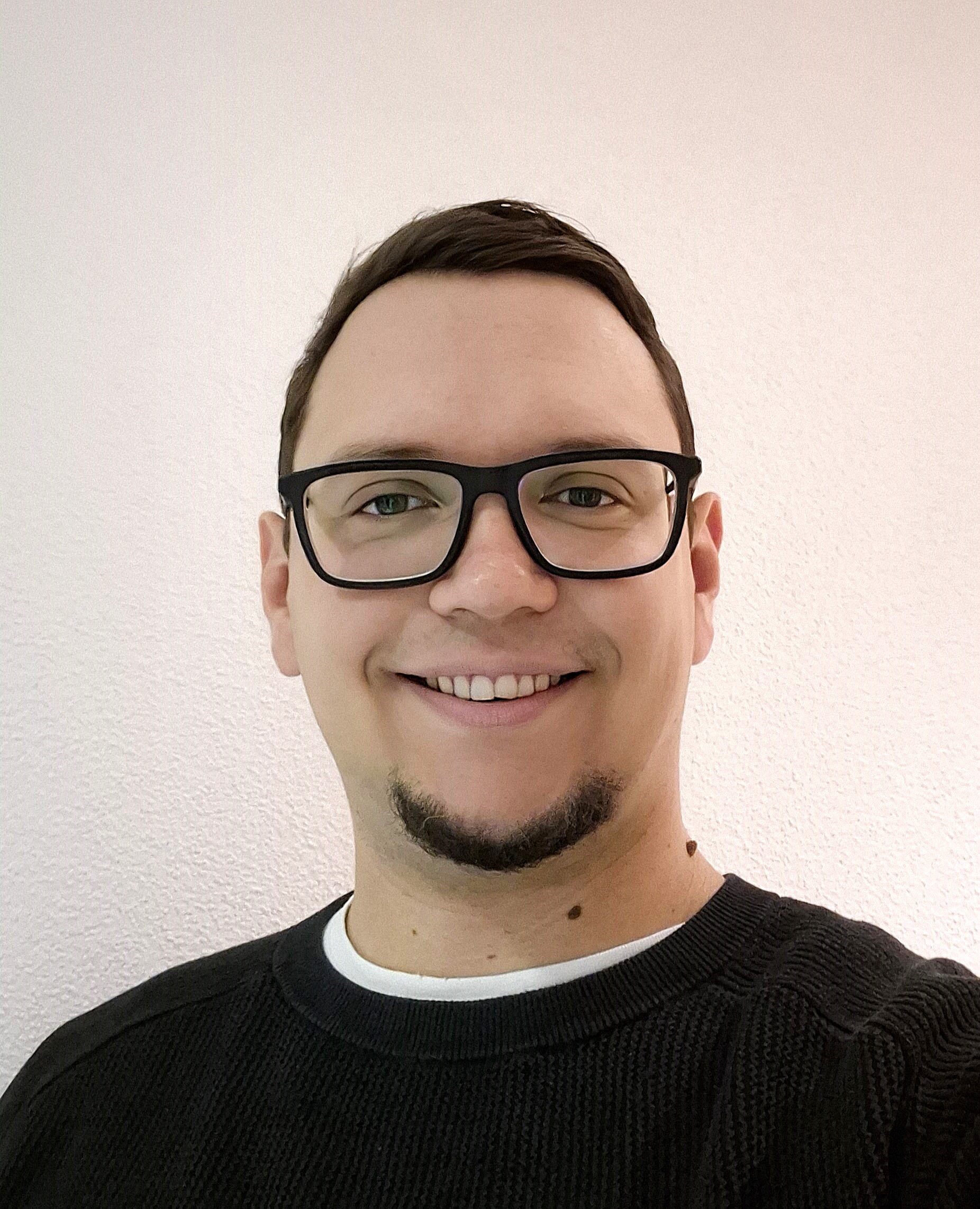}}]{Lukas Schulthess} (GS’22) received the Swiss Certificate of Competence (EFZ) as an electronic technician in 2014. He received the B.Sc. and the M.Sc. degrees in electronics engineering and information technology from ETH Zürich, Switzerland, in 2020 and 2021, respectively. He is currently pursuing a Ph.D. degree at ETH Zürich, Switzerland. His research interests include ultra-low power and miniaturized self-sustainable sensor nodes, wireless body area networks, and energy harvesting.
\end{IEEEbiography}
\vfill\newpage
\begin{IEEEbiography}[{\includegraphics[width=1in,height=1.25in,clip,keepaspectratio]{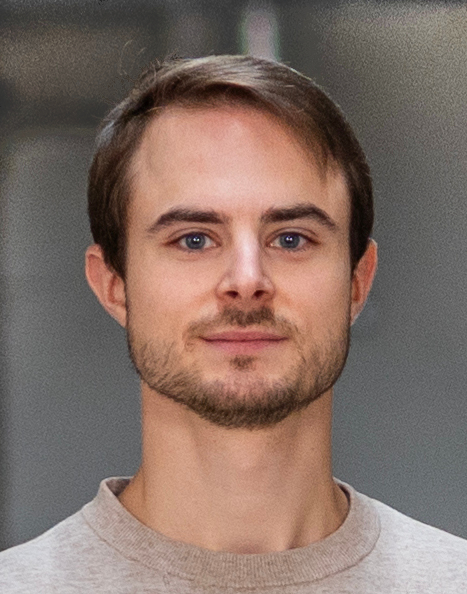}}]{Davide Plozza} (GS’24) received the B.Sc. degree in electronics engineering and information technology in 2020 and the M.Sc. degree in robotics, systems and control in 2023, both from ETH Zürich, Switzerland. He is currently pursuing his Ph.D. degree with the Center for Project-Based Learning at ETH Zürich, Switzerland. His research focuses on robust navigation and locomotion for resource-constrained quadrupedal robots operating in harsh environments.
\end{IEEEbiography}

\begin{IEEEbiography}[{\includegraphics[width=1in,height=1.25in,clip,keepaspectratio]{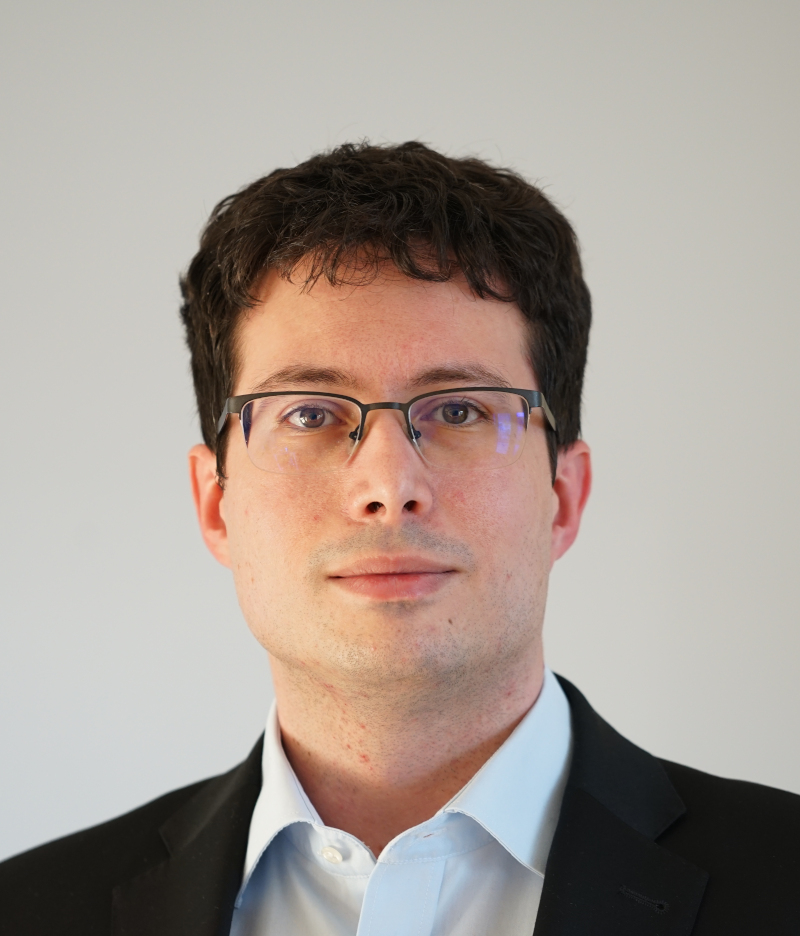}}]{Christian Vogt} (M'20) received the M.Sc. degree and the Ph.D. in electrical engineering and information technology from ETH Zürich, Zürich, Switzerland, in 2013 and 2017, respectively. He is currently a post-doctoral researcher and lecturer at ETH Zürich, Zürich, Switzerland. His research work focuses on signal processing for low power applications, including field programmable gate arrays (FPGAs), IoT, wearables and autonomous unmanned vehicles.
\end{IEEEbiography}

\begin{IEEEbiography}[{\includegraphics[width=1in,height=1.25in,clip,keepaspectratio]{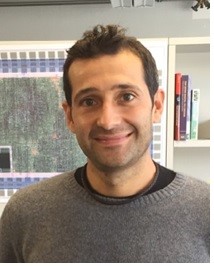}}]{Michele Magno} (SM'13) is currently a Senior Scientist at ETH Zürich, Switzerland, at the Department of Information Technology and Electrical Engineering. Since 2020, he is leading the D-ITET center for project-based learning at ETH. He received his master's and Ph.D. degrees in electronic engineering from the University of Bologna, Italy, in 2004 and 2010, respectively. He is working at ETH since 2013 and has become a visiting lecturer or professor at the University of Nice Sophia, Enssat Lannion, Univerisity of Bologna, and Mid Sweden University. His current research interests include smart sensing, low-power machine learning, wireless sensor networks, wearable devices, energy harvesting, and low-power management techniques. He has authored over 280 publications in international journals and conferences, earning best paper awards at IEEE events as well as recognition for industrial projects and patents.
\end{IEEEbiography}
\vfill
\end{document}